\documentclass{aa}  
\newcommand{\beq}[1]{\begin{equation}\label{#1}}
\newcommand{\eeq}{\end{equation}}

\newcommand{\ms}[1]{\textcolor{black}{ #1}}

\newcommand{\bea}{\begin{eqnarray}}
\newcommand{\eea}{\end{eqnarray}}

\usepackage{orcidlink}

\usepackage{natbib}

\usepackage{url}
\usepackage{soul}
\usepackage{multirow} 
\usepackage{makecell} 
\usepackage{booktabs}
\usepackage{caption}
\usepackage{placeins}
\usepackage{float}
\usepackage{txfonts}
\usepackage{graphicx}
\usepackage{txfonts,textcomp}

\usepackage{natbib,twoopt}
\bibpunct{(}{)}{;}{a}{}{,}
\usepackage[hyphenbreaks]{breakurl}
\usepackage{hyperref}
\hypersetup{
  colorlinks,
  citecolor=cyan,
  linkcolor=magenta,
  urlcolor=teal,
}

\begin{document} 

\title{Binary tidal evolution as a sculptor of circumbinary planet architectures}

\author{
Mario Sucerquia\inst{1}\thanks{mario.sucerquia@univ-grenoble-alpes.fr}\orcidlink{0000-0002-8065-4199}
\and
Nicolás Cuello\inst{1}\email{nicolas.cuello@univ-grenoble-alpes.fr}\orcidlink{0000-0003-3713-8073}
\and
Gaspard Duchêne\inst{1}\email{gaspard.duchene@univ-grenoble-alpes.fr}\orcidlink{0000-0002-5092-6464}
}

\institute{\centering
Univ. Grenoble Alpes, CNRS, IPAG, 38000 Grenoble, France
}

\date{\today}
\abstract
{
Circumbinary planets (CBPs) are expected to form in circumbinary discs and migrate towards the disc inner cavity, often close to the dynamical stability boundary of the binary. Yet, the observed transiting and radial-velocity population is not strictly marginally stable. Excluding extreme systems, the typical offset from the stability limit is $\Delta \equiv a_{\rm p}/a_{\rm crit}=1.22^{+0.18}_{-0.07}$, suggesting that post-formation evolution may have shaped the present-day architecture.
}
{
We investigated whether the secular evolution of close binaries can contribute to the observed displacement of CBPs from marginal stability. We tested how prescribed binary contraction and eccentricity damping move the stability boundary, $a_{\rm crit}(t)$, thereby modifying $\Delta=a_{\rm p}/a_{\rm crit}$ as well as the orbital response, survival, and loss of near-boundary planets as a function of binary mass ratio.
}
{
We performed controlled, high-accuracy $N$-body simulations of single coplanar CBPs initially placed near $a_{\rm crit}$. The planets evolve under Newtonian gravity, while the central binary follows a prescribed contraction and eccentricity damping track that approximately conserves orbital angular momentum. We measured how the resulting motion of $a_{\rm crit}(t)$ modifies the planets' final offsets, orbital response, survival, and loss.
}
{
The retreat of $a_{\rm crit}(t)$ filters the initially near-boundary population, adding a contraction-driven loss component dominated by ejections. Surviving planets remain quasi-fossilised, with median changes in $a_{\rm p}$ and $e_{\rm p}$ close to zero and typical semi-major axis drifts of only a few per cent. Considering all valid synthetic survivors, the final offset reaches a global median value of $\Delta_f = 1.86^{+0.18}_{-0.26}$. Applying the same fiducial damping prescription to the observed sample shifts the median offset from $\Delta_{\rm now}=1.22^{+0.18}_{-0.07}$ to $\Delta_{\rm damp}=1.44^{+0.32}_{-0.10}$. Near-equal-mass binaries undergo stronger dynamical filtering, with enhanced ejections and collisions.
}
{
Binary tidal evolution offers a plausible internal pathway for shaping non-marginal circumbinary architectures. By displacing or removing planets initially close to marginal stability, the inward retreat of $a_{\rm crit}(t)$ can widen the inner circumbinary gap and move surviving planets towards larger offsets from the present-day stability boundary. This mechanism links binary orbital evolution, dynamical filtering, and the detectability of transiting CBPs, supporting a multi-channel interpretation of the CBP desert in which physical sculpting and geometric selection act together.
}


\titlerunning{Binary tidal evolution as a sculptor of circumbinary planet architectures}
\authorrunning{Sucerquia, Cuello \& Duchêne}
\maketitle
\nolinenumbers
%

\section{Introduction}\label{sec:introduction}

Surveys combining {Gaia} astrometry, high-resolution imaging, and long-baseline radial velocities show that roughly 40--50\% of solar-type stars reside in multiple systems, with the multiplicity fraction increasing towards higher stellar masses \citep{Duchene2013,Offner2023}. Many of these systems are hierarchical binaries with well-separated dynamical scales, allowing circumstellar or circumbinary discs to form and evolve over long timescales \citep{Tokovinin2014a,Tokovinin2014b}. They are therefore common and dynamically diverse sites of planet formation.

Planets in binary systems are now observed across several architectures. 
{Kepler}, TESS, BEBOP, and microlensing programmes have revealed planets orbiting one or both components of stellar binaries, and PLATO is expected to expand this census \citep[e.g.][]{Doyle2011,Welsh2012,Kostov2020,Standing2022,Bennett2016,Rauer2014}. 
Unlike planets around single stars, their orbits are constrained by the binary potential: stability boundaries delimit the permitted regions \citep{Dvorak1989, Holman_Wiegert1999,Pilat-Lohinger2003, Quintana2006, Trani2022,Georgakarakos2024}, while forced eccentricities and precession shape the surviving architectures \citep{Leung2013}.

Planets in binary systems fall into two broad dynamical categories \citep{Haghighipour2006}. 
S-type planets orbit one stellar component. When the hierarchy is strong, the companion acts mainly as a secular perturber and long-term stability is common \citep{Hamers2016}. 
This configuration accounts for most confirmed planets in binaries ($\sim96$\%; \citealt{Thebault2025}), and its architectures often remain close to those around single stars \citep{Hamers2015}.

Circumbinary, or P-type, planets by contrast orbit the barycentre of the stellar pair and sample a time-dependent, non-Keplerian potential. Their forced eccentricities, precession, and stable regions depend on the binary semi-major axis, eccentricity, and mass ratio $q_{\rm B}\equiv m_2/m_1$ (with $m_2\le m_1$; see e.g. \citealt{Leung2013}). The perturbation is strongest near equal masses, where the non-axisymmetric component of the binary potential is largest. As a result, long-lived circumbinary planets (CBPs) are confined to orbits outside a critical stability boundary, $a_{\rm crit}$ \citep{Holman_Wiegert1999,Trani2022,Georgakarakos2024}.

The stability of P-type orbits has been extensively studied for fixed binary configurations. In this static problem, the location of $a_{\rm crit}$ is controlled by the binary semi-major axis, eccentricity, and mass ratio, together with the planet's eccentricity, inclination, and resonant environment \citep{Dvorak1989,Holman_Wiegert1999,Pilat-Lohinger2003,Quintana2006}. 
This stability constraint leaves long-lived CBPs on relatively wide orbits, with long periods, reduced transit probabilities, and strong transit-time variations \citep{Martin2014}. 
The observed sample is therefore shaped by both dynamical filtering and observational selection and should be treated as a biased but informative view of the underlying circumbinary population \citep{Welsh2012}.

\begin{figure} \centering %

\includegraphics[width=0.48\textwidth]{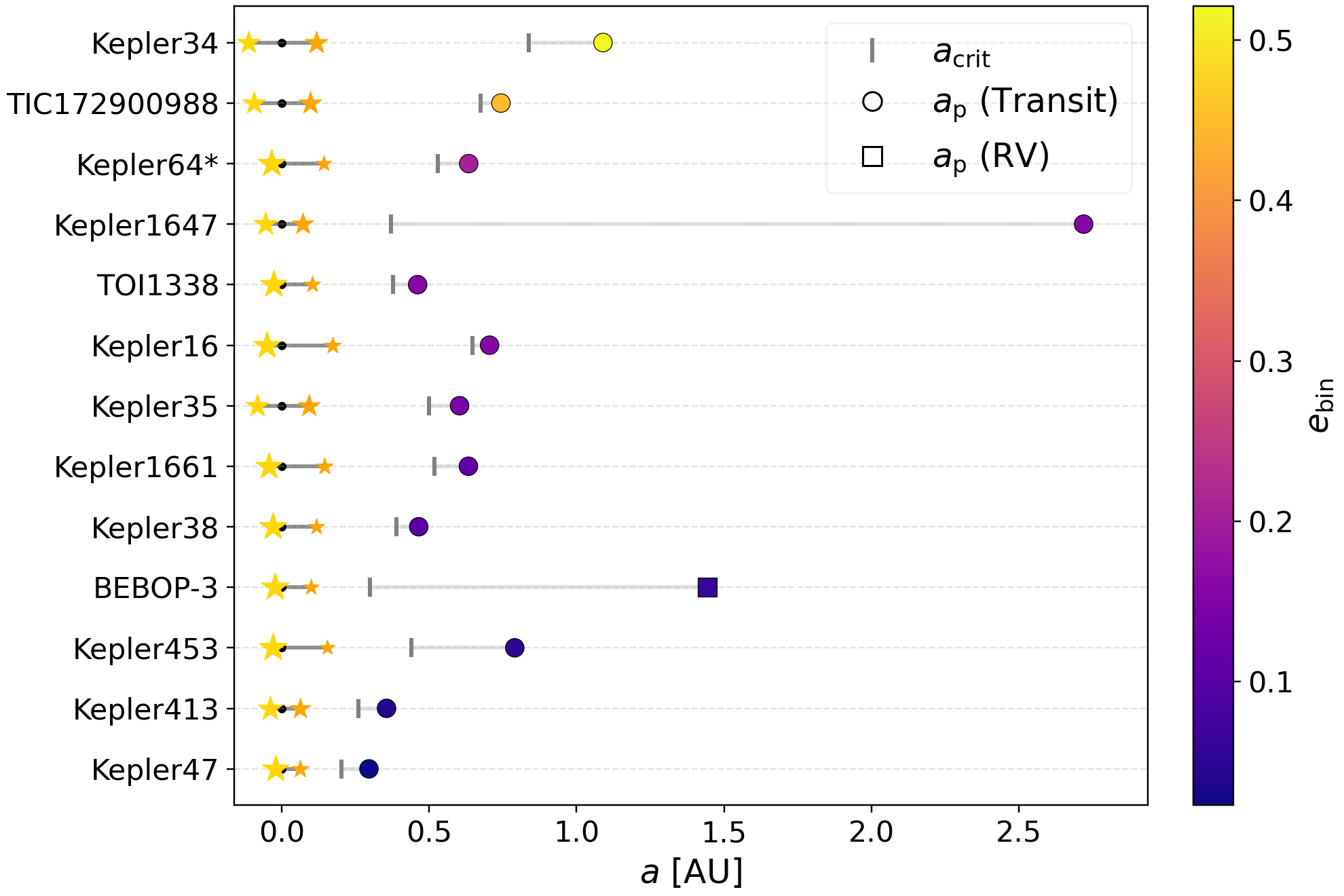} \caption{Semi-major axes of the confirmed CBPs in our sample compared with the corresponding dynamical stability limits, $a_{\rm crit}$ (grey markers). The circular symbols denote planets detected by transit, while the square marks the system detected by radial velocity. Planetary semi-major axes are colour-coded by the binary eccentricity, $e_{\rm bin}$, and the systems are ordered from top to bottom by decreasing $e_{\rm bin}$. For each row, the two star symbols indicate the schematic positions of the binary components relative to the barycentre (black dot), using the binary semi-major axis and stellar mass ratio. The horizontal axis is shown on a linear scale to emphasise the relative location of each planet with respect to the stability boundary.}\label{fig:populationgaps} 
\end{figure}

\begin{figure} \centering %
\includegraphics[width=0.48\textwidth]{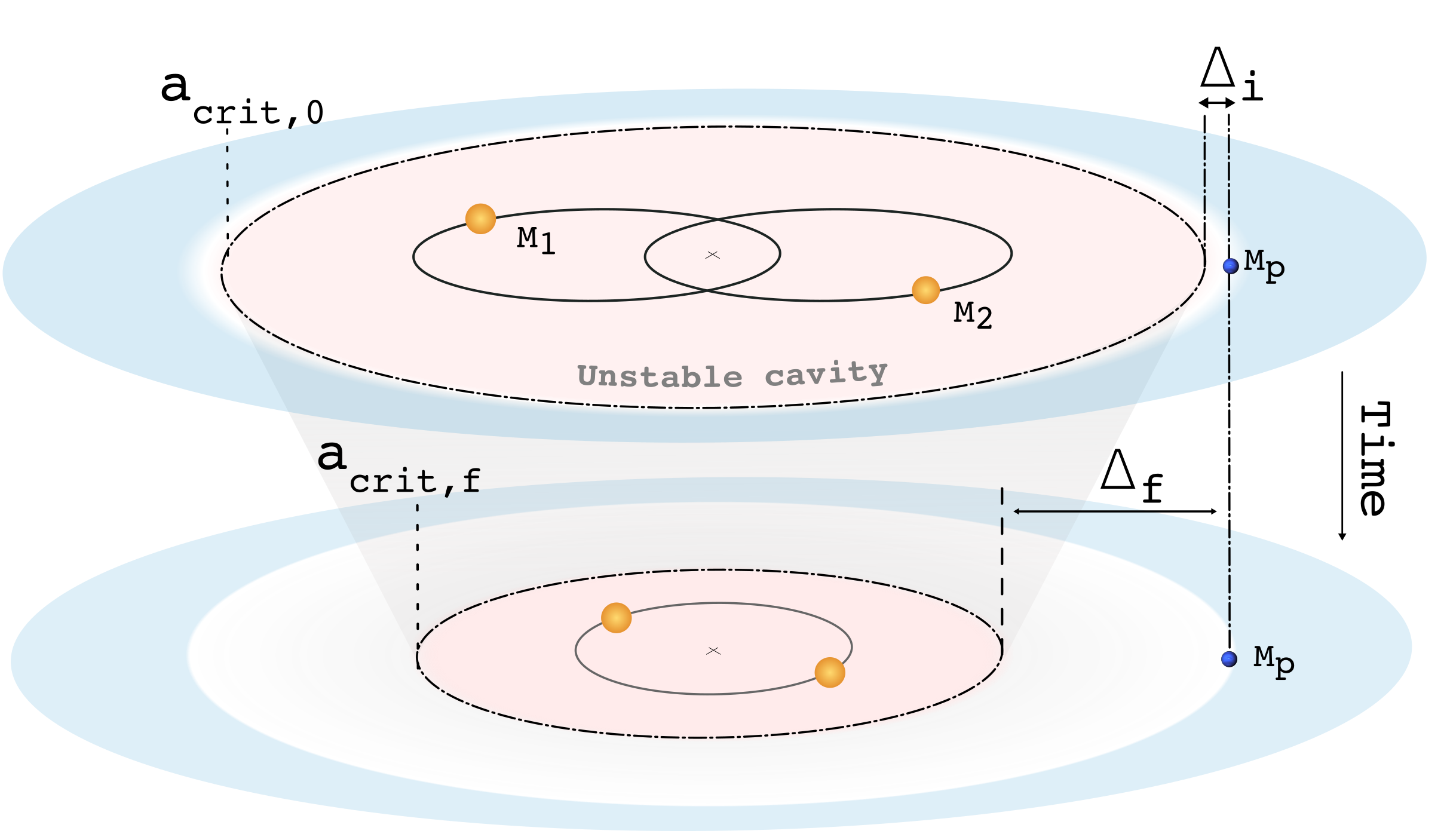} \caption{ Schematic illustration of the circumbinary stability boundary before (top) and
after (bottom) tidal contraction and eccentricity damping of the binary.
The red region indicates the dynamically unstable circumbinary cavity, while the
blue region corresponds to long-term stable orbits.
The critical stability radius shifts from $a_{\rm crit,0}$ to $a_{\rm crit,f}$,
changing the effective offset between the planet ($M_{\rm p}$) and the stability
boundary from $\Delta_i$ to $\Delta_f$. } \label{fig:mechanism} 
\end{figure}

We use this static framework as a reference but consider a different problem: a close binary whose orbit changes with time, carrying the stability boundary with it. During the late disc phase, CBPs embedded in gaseous discs are expected to migrate inwards towards the disc inner cavity, where the net disc torque vanishes \citep{Pierens2007,Pierens2008,Mutter2017,Thun2018}. The parking radius need not coincide exactly with the dynamical stability boundary, but it can lie close to it, depending on the binary--disc interaction \citep{Thun2017}. A primordial circumbinary population should therefore emerge from the disc phase at separations not far from $a_{\rm crit}$. The observed compact CBP population is displaced from this near-boundary picture. The innermost known transiting and radial-velocity planets lie modestly beyond the stability boundary rather than at strict marginal stability (Fig.~\ref{fig:populationgaps}). We discuss quantifying this displacement in Sect.~\ref{sec:demographic} and use it as a reference scale for our dynamical experiments.

\begin{figure*}
  \centering
  \includegraphics[width=0.95\columnwidth]{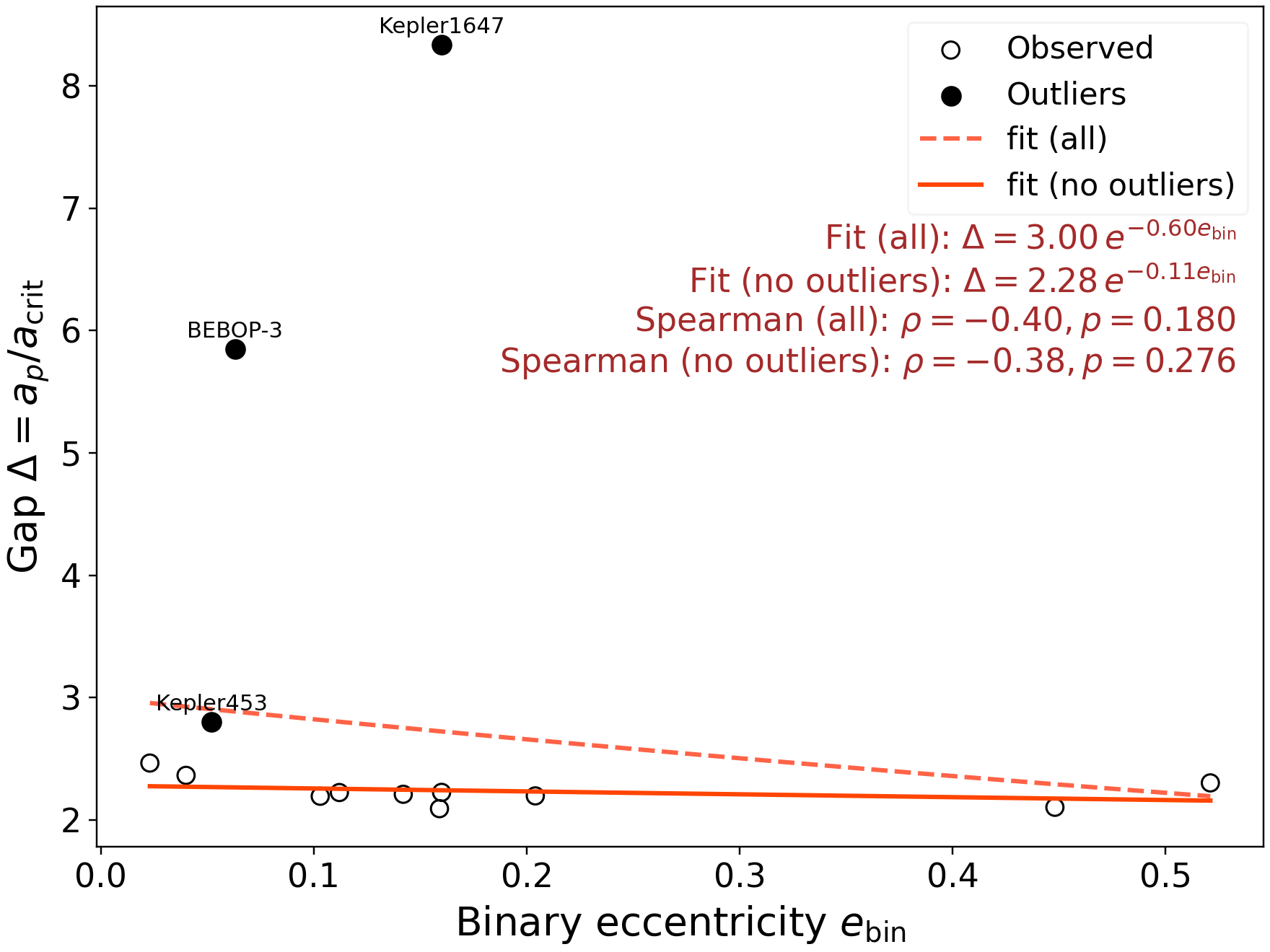}
  \includegraphics[width=0.95\columnwidth]{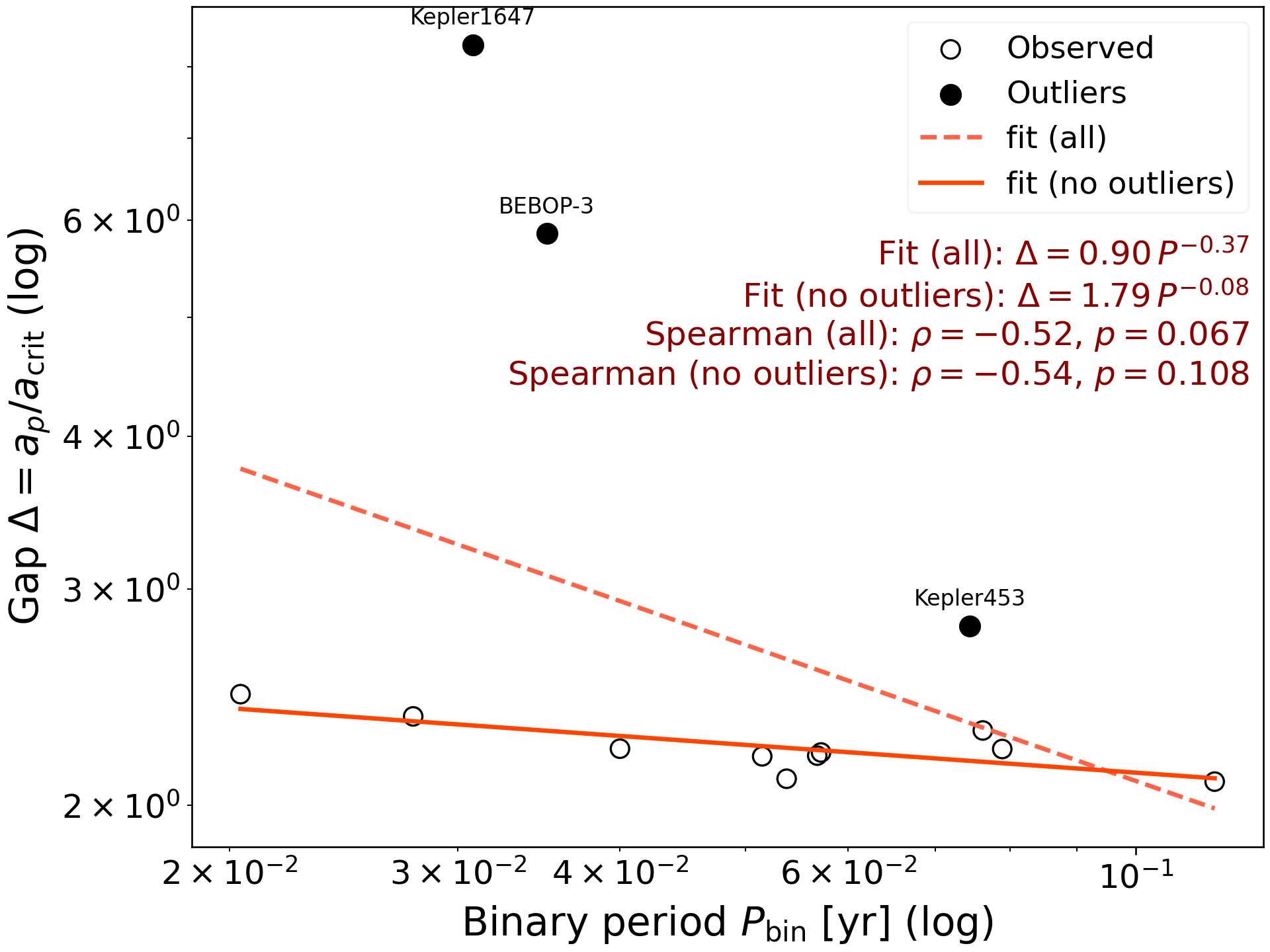}
  \caption{
  Left panel:
  Offset from the dynamical stability limit,
  \(\Delta = a_p/a_{\rm crit}\), as a function of the binary eccentricity
  \(e_{\rm bin}\), for the confirmed transiting and radial-velocity CBPs.
  The open circles denote systems consistent with the main circumbinary population,
  while the filled circles highlight outliers (Kepler--1647, BEBOP--3, Kepler--453)
  whose architectures likely reflect distinct evolutionary pathways.
  The dashed line shows a linear fit to all systems in
  \(\ln\Delta\)–\(e_{\rm bin}\) space, and the solid line to the sample with
  the outliers removed.
  Right panel:
  Same offset \(\Delta\) plotted against the binary period \(P_{\rm bin}\)
  on logarithmic axes. The dashed and solid lines show linear fits in
  \(\ln\Delta\)–\(\ln P_{\rm bin}\) space for the full sample and the
  non-outlier subsample, respectively. 
  }
  \label{fig:tendencies_Delta_a}
\end{figure*}

Several processes can move CBPs away from strict marginal stability. 
Early binary--disc interactions can change the shape and radial extent of the circumbinary cavity, while instabilities during multi-planet migration may clear the innermost stable orbits and leave the surviving planets at larger separations \citep{Grane2026}. 
The binary can also evolve after the disc phase. If tidal dissipation contracts and circularises the stellar orbit, the stability boundary $a_{\rm crit}(t)$ retreats inward, widening the planet--boundary separation even when the planetary semi-major axis changes little. 
Although this mechanism follows directly from the dependence of $a_{\rm crit}$ on the binary orbit, its population-level consequences have not been quantified.

Close stellar binaries are not static backgrounds for their CBPs. 
Tidal dissipation within the stars can shrink the binary orbit and damp its eccentricity, with an efficiency that depends on stellar structure and evolutionary stage \citep{Sun2018,Terquem2021}. 
This expectation is supported observationally by close binaries with evolved red-giant primaries, which exhibit lower eccentricities at fixed orbital period than less evolved systems \citep{Beck2024}.

The binary mass ratio is also relevant. 
For the close solar-type binaries considered here, observed samples are weighted towards comparable-mass companions, with $q_{\rm B}\!\gtrsim\!0.5$ common at $P_{\rm bin}\!\lesssim\!100$~d \citep{Raghavan2010}, although low-$q_{\rm B}$ systems may be undercounted in spectroscopic surveys \citep{Moe2017}. 
Very low mass ratios are further suppressed around low-mass primaries by the paucity of brown-dwarf companions \citep{Winters2019}.

In this work, we treated close solar-type binaries as tidally evolving systems and followed prescribed contraction and eccentricity damping tracks in the $(a_{\rm bin}, e_{\rm bin})$ plane. 
As the binary evolves, $a_{\rm crit}(t)$ changes with it, modifying the distance between the planet and the stability boundary even if the planetary orbit remains nearly fixed (Fig.~\ref{fig:mechanism}). 
We used controlled $N$-body experiments to quantify this effect for initially near-boundary CBPs, measuring their orbital response, survival, ejection, and Roche-loss rates as a function of binary mass ratio. In Sect.~\ref{sec:demographic} we place the current population of CBPs in context. Section~\ref{sec:methodology} presents our model for binary tidal shrinkage and eccentricity damping. The main dynamical outcomes of our simulations are detailed in Sect.~\ref{sec:results}. Finally, Sect.~\ref{sec:discussion} discusses the implications of our findings, and Sect.~\ref{sec:conclusions} summarises our main conclusions.

\section{Circumbinary planet demographics}\label{sec:demographic}

The confirmed CBP sample spans a heterogeneous set of host binaries and discovery channels. In the current \texttt{exoplanet.eu} circumbinary compilation, there are 35 systems: 12 discovered via transits, 1 via radial velocities, and 22 via other techniques (timing, microlensing, astrometry, and imaging). 
This heterogeneity matters for demographics: transit and radial-velocity (RV) surveys preferentially probe compact, well-characterised binaries (often eclipsing) hosting close-in planets, whereas the `other' category is dominated by systems where the binary or planetary architecture is much wider, dynamically different (e.g. post-common-envelope timing systems), or characterised by less homogeneous orbital information, making direct comparisons in terms of $a_{\rm crit}$ less straightforward.
Because our goal is to characterise the architecture of close CBPs relative to a well-defined dynamical stability limit, we focus below on the subset of confirmed transiting and RV CBPs for which the binary orbit is well constrained and a homogeneous computation of $a_{\rm crit}$ is meaningful (Fig.~\ref{fig:tendencies_Delta_a}).

A key population-level clue comes from the fractional separation between each planet and the dynamical stability boundary, quantified through the offset
$\Delta \equiv a_{\rm p}/a_{\rm crit}$.
In a homogeneous compilation of transiting and RV CBPs, the population avoids strict marginal stability: excluding extreme systems, the typical separation is
$\Delta = 1.22^{+0.18}_{-0.07}$ (16--84\% range), rather than clustering near the stability limit at $\Delta \simeq 1$ (Fig.~\ref{fig:tendencies_Delta_a}).
This offset is unlikely to be produced by observational selection alone, since detectability generally favours short-period planets (and therefore smaller $a_{\rm p}$), which would tend to decrease $\Delta$ rather than systematically increase it.

We note that $\Delta$ is computed for the `innermost detected' planet in each system, which may not always coincide with the true innermost planet if additional closer-in companions remain undetected. Such incompleteness would bias the observed $\Delta$ distribution towards larger values, potentially strengthening the conclusion that many CBPs reside close to marginal stability. This caveat is particularly relevant when combining transit-selected systems with RV surveys such as BEBOP \citep[e.g.][]{Standing2023}.

When examined against binary properties, the observed offsets show suggestive structure, albeit in a small sample. In Fig.~\ref{fig:tendencies_Delta_a} we fit $\ln \Delta$ as a function of $e_{\rm bin}$ and $\ln P_{\rm bin}$ for (i) the full sample and (ii) a subsample excluding the most extreme architectures (Kepler--1647, BEBOP--3, Kepler--453). The largest values of $\Delta$ appear preferentially in the most compact binaries, whereas any trend with present-day binary eccentricity remains weak and should be regarded as tentative. Given the small number of systems and the dominance of the transiting Kepler/TESS discoveries, we treat these patterns as indicative rather than definitive.

The outliers occupy a distinct region of parameter space. Kepler--1647 and BEBOP--3 reside at unusually large separations from the stability boundary, while Kepler--453 is known to be dynamically misaligned with respect to its host binary. These systems may therefore reflect additional degrees of freedom not captured by the main, near-coplanar transiting population (e.g. distinct formation environments, dynamical excitation, or unseen companions), and we do not attempt to interpret them with the same minimalist framework used for the bulk population.

Altogether, the observed CBP demographics provide a clear target for theory: any successful mechanism must explain why the typical system sits at $\Delta\simeq 1.2$ (instead of $\Delta\simeq 1$), and why the largest offsets appear preferentially in the most compact binaries. 
In the next sections, we discuss developing a dynamical model in which the binary itself evolves secularly (through tidal shrinkage and eccentricity damping), thereby moving $a_{\rm crit}(t)$ and potentially imprinting the observed offsets without requiring additional post-formation planetary migration.

\section{Methodology}
\label{sec:methodology}

 {We focus on close solar-type binaries with separations $a_{\rm bin}\lesssim0.3$ AU. In classical stellar-dynamical terminology, these systems are `hard' binaries, whose binding energies exceed the kinetic energy of typical field stars \citep{Heggie1975,BinneyTremaine2008}. They are therefore effectively immune to disruption by stellar encounters, allowing us to treat them as isolated systems whose long-term orbital evolution is governed primarily by internal processes, in particular stellar tides. In the present work, this evolution is prescribed, and the integrations are kept purely Newtonian to isolate the response of near-boundary CBPs to the imposed retreat of $a_{\rm crit}(t)$. We therefore neglect relativistic apsidal precession and do not follow the secular-resonance channel described by \citet{Farhat2025}, in which relativistic precession of the binary can couple to planetary apsidal precession during binary inspiral. This omission is part of the controlled design of the experiment, but it also represents an important limitation for the most compact binaries.}

\subsection{Dynamical evolution}
\label{sec:bin_evol}

Our goal is to quantify how a CBP reacts when the central binary undergoes a slow and secular orbital evolution. We performed controlled numerical experiments with the \texttt{REBOUND} $N$-body package \citep{rebound} using the high-order integrator \texttt{IAS15} \citep{reboundias15}. Each simulation contained a stellar binary and a single coplanar CBP interacting only through Newtonian gravity. The binary evolution is not produced by a physical tidal model inside the code. Rather, it is imposed externally through a parametric prescription. This choice allows us to isolate the dynamical imprint of binary eccentricity damping from the uncertainties of stellar tidal dissipation.

\subsubsection{Parametric evolution of the inner binary}

We adopted an idealised model in which the binary eccentricity decays exponentially during a prescribed interval $[T_0,\,T_0+\Delta T]$:
\begin{equation}
    e_{\mathrm{bin}}(t)
    = \left(e_0-e_{\mathrm{fin}}\right)\exp\!\left[-\frac{t-T_0}{\tau_e}\right] + e_{\mathrm{fin}} ,
\end{equation}
so that $e_{\mathrm{bin}}(t)$ decreases smoothly from $e_0$ to a residual value $e_{\mathrm{fin}}$ on a timescale $\tau_e$.
At the same time, we enforced approximate conservation of the binary orbital angular momentum by keeping
\begin{equation}
    a_{\mathrm{bin}}(t)\,\left[1 - e_{\mathrm{bin}}^2(t)\right]
    = a_0\left(1-e_0^2\right)
    \equiv \mathrm{const}.
\end{equation}
This prescription dissipates orbital energy while retaining angular momentum within the binary, capturing the essential secular behaviour expected during eccentricity damping and partial circularisation while remaining agnostic about the tidal microphysics.

\subsubsection{Numerical realisation of the forcing}

To realise the prescribed time-dependence $(a_{\rm bin}(t),e_{\rm bin}(t))$ within an $N$-body integration, we applied small velocity corrections to the `relative' motion of the stars during the forcing interval $[T_0,\,T_0+\Delta T]$. At each step we worked in the instantaneous relative frame, with $\mathbf{r}=\mathbf{r}_2-\mathbf{r}_1$, $\mathbf{v}=\mathbf{v}_2-\mathbf{v}_1$, and $\hat{\mathbf{r}}=\mathbf{r}/r$. We decomposed the relative velocity into radial and tangential parts,
\begin{equation}
v_r = \mathbf{v}\cdot \hat{\mathbf{r}}, \qquad 
\mathbf{v}_t = \mathbf{v} - v_r\,\hat{\mathbf{r}},
\end{equation}
so that the instantaneous specific angular momentum is $h=|\mathbf{r}\times\mathbf{v}|=r\,v_t$.

This forcing was implemented as a non-conservative `orbit-modification' operator applied to the inner binary, following the spirit of synthetic prescriptions commonly used in $N$-body experiments (e.g. \texttt{REBOUNDx}; \citealt{reboundx}). The operator mimics tidal circularisation by dissipating orbital energy while conserving the binary angular momentum to first order. We therefore kept $\mathbf{v}_t$ unchanged (thus preserving $h$ at fixed $r$) and updated only the radial component so that the instantaneous specific orbital energy matched the target semi-major axis $a_{\rm bin}(t)$ through the vis-viva relation,
\begin{equation}
\frac{1}{2}\left(v_t^2+v_r'^2\right) - \frac{G M_{\rm tot}}{r}
\;=\; -\frac{G M_{\rm tot}}{2a_{\rm bin}(t)},
\end{equation}
with the additional convention ${\rm sgn}(v_r')={\rm sgn}(v_r)$ to preserve the direction of motion along $\hat{\mathbf{r}}$.
Equivalently, the required radial speed is obtained from
\begin{equation}
v_r'^2 \;=\; \max\!\left[0,\; G M_{\rm tot}\!\left(\frac{2}{r}-\frac{1}{a_{\rm bin}(t)}\right) - v_t^2 \right],
\end{equation}
where the $\max$ prevents spurious negative values near turning points. The corrected relative velocity is then $\mathbf{v}'=\mathbf{v}_t+v_r'\,\hat{\mathbf{r}}$.

For numerical smoothness we capped the correction by limiting $|v_r'-v_r|$ to a small fraction of the local orbital speed, ensuring that the imposed evolution remains adiabatic relative to the binary period. Finally, the relative impulse $\Delta\mathbf{v}=\mathbf{v}'-\mathbf{v}$ was distributed between the stars according to their masses,
\begin{equation}
\Delta\mathbf{v}_1 = -\frac{m_2}{M_{\rm tot}}\,\Delta\mathbf{v}, \qquad
\Delta\mathbf{v}_2 = +\frac{m_1}{M_{\rm tot}}\,\Delta\mathbf{v},
\end{equation}
which keeps the barycentric velocity unchanged and therefore preserves the system centre of mass.

\subsubsection{Integration procedure and orbital diagnostics}

Simulations were performed using \texttt{IAS15} with adaptive time steps, adopting years, astronomical units, and solar masses as the units of time, length, and mass, respectively. Each realisation began at $t=0$ and was evolved until
\begin{equation}
    T_{\mathrm{end}} = T_0 + \Delta T + \Delta T_{\mathrm{post}},
\end{equation}
including a post-eccentricity damping phase where the binary was held fixed.

Unless otherwise stated, we adopted a fiducial binary evolution with $T_0=\Delta T= \Delta T_{\rm post}=100~{\rm kyr}$, so that $T_{\rm end}=300~{\rm kyr}$. The eccentricity damping time was $\tau_e=100~{\rm kyr}$, with $e_0=0.40$ and $e_{\rm fin}=0.05$; therefore, the binary reached $e_{\rm bin}\simeq0.18$ by the end of the forcing interval. Thus, the fiducial simulations correspond to binary contraction and partial circularisation rather than complete circularisation. These timescales are not intended to represent a specific stellar tidal model. They were chosen to make the imposed binary evolution secular with respect to the orbital periods, while allowing us to isolate the dynamical response of near-boundary CBPs to a controlled retreat of $a_{\rm crit}(t)$.  {To test the sensitivity to this timescale, we repeated a reduced $12\times12$ subset of the initial-condition grid for $\tau_e=\Delta T=30$ and $300$ kyr, and compared these runs with the fiducial $100$ kyr case. We considered $q_{\rm B}=0.30$, $0.70$, and $1.00$, while keeping the binary evolutionary track and all other parameters unchanged. The results of this test are presented in Appendix~\ref{app:timescale}.}

The circumbinary orbit exhibits fast oscillations at harmonics of the binary period, which can obscure secular trends in $a_p$ and $e_p$. To suppress this `breathing', we recorded planetary orbital elements through a stroboscopic sampling tied to the binary phase and applied a short moving average to the stroboscopic sequence. From these filtered samples we defined two phase-averaged states: an initial state measured immediately before the forcing begins, $(a_i,e_i)$ and a final state measured after the forcing phase, $(a_f,e_f)$. 
These define our main dynamical diagnostics,
\begin{equation}
    \Delta a = a_f - a_i, \qquad \Delta e = e_f - e_i,
\end{equation}
as well as the instantaneous separations from the moving stability boundary. At each stroboscopic sample, we computed $a_{\rm crit}(t)$ from the binary elements, and we evaluated
\begin{equation}
    \Delta(T_0)=\frac{a_i}{a_{{\rm crit},i}},  \quad
    \Delta_f=\frac{a_f}{a_{{\rm crit},f}} .
\end{equation}
This construction is central for connecting the simulations with the observed CBP offsets, since it measures the gap relative to the stability limit at well-defined evolutionary stages rather than by initial design.

\subsection{Exploration of parameter space and phase randomisation}
\label{sec:parspaceexpl}

We explored a 2D grid in
\begin{equation}
    (f,\,e_{p,0}) = \left(\frac{a_{p,0}}{a_{\mathrm{crit},0}},\, e_{p,0}\right),
\end{equation}
where $a_{\mathrm{crit},0}$ is evaluated from the initial binary elements. For each grid point, we ran a pair of simulations: (i) a control integration with no tidal evolution and (ii) an otherwise identical integration where the binary follows the imposed contraction and eccentricity-damping track. To reduce phase-driven variance and enable one-to-one comparisons, we randomised the initial orbital phases but kept them `matched' between the paired runs. We drew the binary and planetary mean anomalies, and, when $e_{p,0}>0$, the planetary argument of pericentre $\omega_p$, uniformly on $[0,2\pi)$ using a deterministic seed assigned to each grid point.

The explored binary mass-ratio range was $q_{\rm B}=\{0.05,0.10,\ldots,1.00\}$, with a spacing $\Delta q_{\rm B}=0.05$. For each $q_{\rm B}$, we sampled a $20\times20$ grid in $(f,e_{p,0})$, with $f\in[0.90,1.50]$ and $e_{p,0}\in[0,0.50]$. The planet mass was kept fixed in all simulations at $m_p=3\times10^{-6}\,M_\odot$, approximately one Earth mass. \ms {We tested the sensitivity to this choice by repeating a reduced grid with a Jupiter-mass planet and a Jupiter-like bulk density. The survival statistics and final offsets of the surviving population remain nearly unchanged, although the relative contribution of ejections and Roche losses shows some sensitivity to the planetary density. These results are presented in Appendix~\ref{app:planet_mass}}.

\subsection{Outcome classification: survival, Roche loss, and ejection}

We stopped integrations at $T_{\rm end}$ or when the planet was lost. A planet `survives' if it remains bound and the integration reaches $T_{\rm end}$.
A `collision' corresponds to a physical loss inside a Roche limit. We assigned each star an effective radius equal to the fluid Roche limit for a rocky planet of density $\rho_p$,
\begin{equation}
    R_{\rm Roche} = k\,R_\star\left(\frac{\rho_\star}{\rho_p}\right)^{1/3},
\end{equation}
with $k=2.44$. Stellar radii were modelled through a main-sequence mass--radius relation $R_\star/R_\odot=(M_\star/M_\odot)^\alpha$ (we adopted $\alpha=0.8$), and $\rho_\star$ is the corresponding mean density. Collisions were detected using REBOUND's direct collision handling, and each integration was halted at the first Roche-crossing event. We recorded which star was impacted.

An ejection was recorded when the planet exceeded a distance
threshold $R_{\rm eject}=10$ AU and exhibited positive specific orbital energy relative to the binary barycentre, using $M_{\rm tot}$ as the effective central mass. We adopted $R_{\rm eject}=10$ AU as an operational escape proxy on the timescales considered.

Although the present study focuses on the dynamical response of a planet to a prescribed tidal evolution of the binary, the adopted initial $(a_{\rm bin},e_{\rm bin})$ are physically motivated. Hydrodynamic simulations show that circumbinary discs can excite significant eccentricities and modify the early binary orbit \citep[e.g.][]{Penzlin2025}, imprinting a pre-processing stage before tidal eccentricity damping becomes dominant. Our initial conditions therefore represent plausible post-disc configurations rather than arbitrary choices.

Our numerical experiments model a `single' CBP per system. Therefore, the results should be interpreted as describing the response and fate of an initially stable planet to the evolving binary potential, without back-reaction from additional planets. Multi-planet interactions (e.g. resonant chains, scattering among planets, or secular coupling between planets) are not included here and are left for future work.

\begin{figure} \centering \includegraphics[width=0.97\columnwidth]{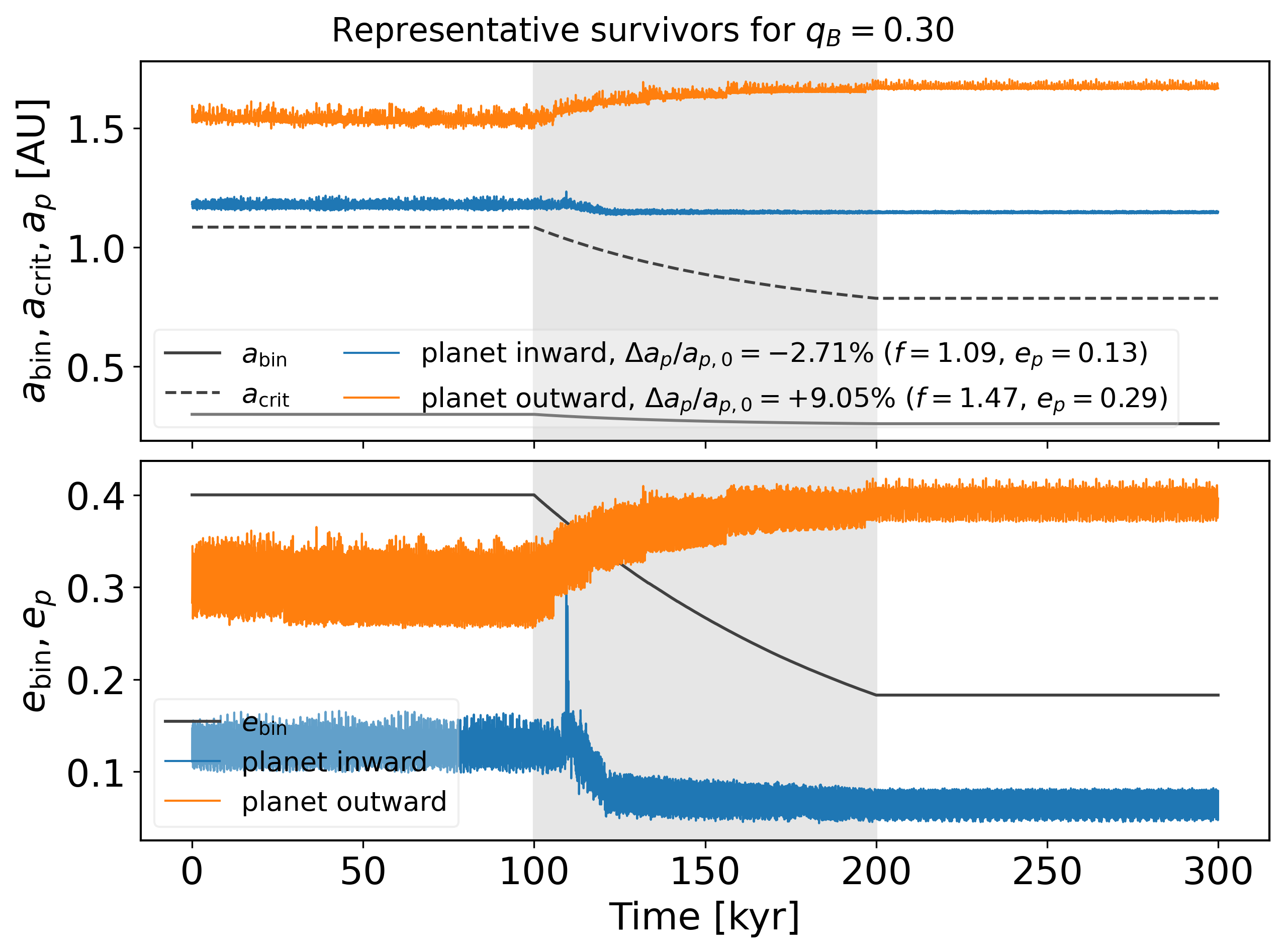} \caption{Example of the orbital response of CBPs during the tidal contraction of their host binary. We show two extreme survivors from the same numerical experiment: one experiences a small inward drift (blue) and the other a small outward drift (orange). The upper panel displays the evolution of the binary semi-major axis $a_{\rm bin}$, the moving stability boundary $a_{\rm crit}$, and the planetary semi-major axes.   {The lower panel shows the evolution of the binary and planetary eccentricities.}} \label{fig:showcase} \end{figure}
\begin{figure*}
\centering
\begin{minipage}[c]{0.7\textwidth}
    \centering
    \includegraphics[width=\linewidth]{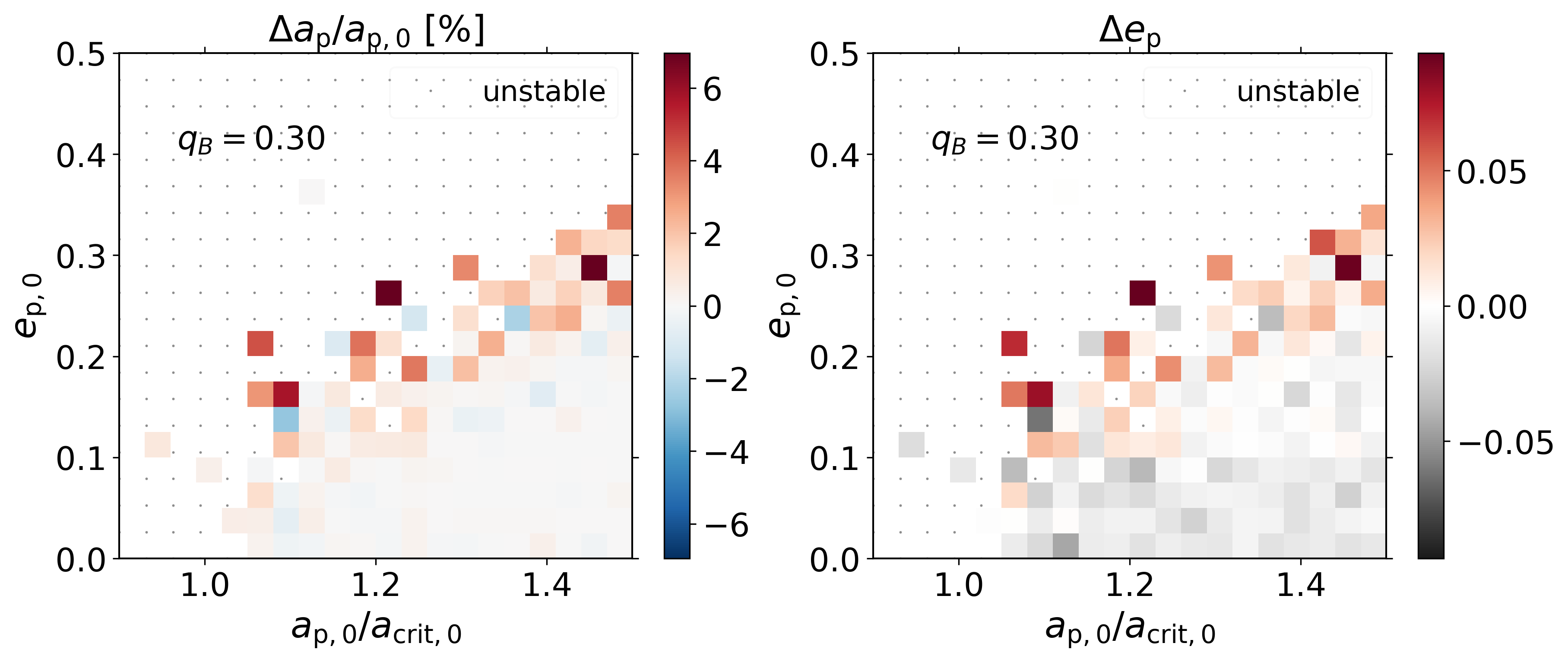}
\end{minipage}
\hfill
\begin{minipage}[c]{0.29\textwidth}
    \captionof{figure}{
    Orbital response across the $(f,e_{p,0})$ grid for a representative binary mass ratio, $q_B=0.30$, where
    $f=a_{p,0}/a_{\rm crit,0}$. Colours show the relative semi-major axis drift, $\Delta a_p/a_{p,0}$, in the left panel, and the eccentricity change, $\Delta e_p$, in the right panel. Grey dots mark unstable outcomes, including ejections and Roche collisions; only surviving planets are colour-coded. The full set of maps for all explored mass
    ratios is shown in Appendix~\ref{app:maps_full}.
}
    \label{fig:cbp_maps_q}
\end{minipage}

\end{figure*}
\section{Results}
\label{sec:results}

\subsection{Orbital evolution in representative examples}
\label{subsec:showcase}

To illustrate the orbital response during binary contraction, Fig.~\ref{fig:showcase} shows two surviving planets selected to bracket the range of semi-major axis responses in the $q_{\rm B}=0.30$ experiment. The shaded interval marks the forcing phase, during which the binary shrinks and circularises from $e_{\rm bin}=0.40$ to $e_{\rm bin}\simeq 0.18$. As a consequence, the stability boundary retreats inwards by $\simeq 27\%$, from $a_{\rm crit}\simeq 1.08$ to $\simeq 0.79$~AU. The two planets are initially located at different fractional separations from the stability limit, $f=a_{p,0}/a_{{\rm crit},0}=1.09$ and $1.47$, and both remain bound throughout the integration.

Despite the substantial displacement of $a_{\rm crit}(t)$, the planetary semi-major axes undergo only modest secular changes. The inner survivor experiences a small inward drift, $\Delta a_p/a_{p,0}=-2.71\%$, whereas the outer survivor drifts outwards by $\Delta a_p/a_{p,0}=+9.05\%$. In both cases, the change in the planet--stability boundary separation is therefore driven mainly by the retreat of $a_{\rm crit}$ rather than by large-scale planetary migration: $\Delta(t)=a_p/a_{\rm crit}(t)$ increases during the forcing phase primarily because $a_{\rm crit}$ moves inward.

The eccentricity response is more pronounced and depends on the trajectory through phase space. While the binary circularises, one of the planets undergoes eccentricity damping (from $\sim0.33$ to $\sim0.23$), whereas the other experiences excitation up to $\sim0.40$. After the forcing ends, both systems settle into nearly stationary configurations with small residual oscillations.

These examples correspond to the largest secular excursions among surviving planets in our survey for $q_{\rm B}=0.30$. Even in these extreme cases, the migration in $a_p$ remains modest compared to the retreat of $a_{\rm crit}(t)$, supporting the idea that CBPs can remain quasi-fossilised while tidal evolution primarily reshapes the inner stability boundary.

\subsection{Orbital response across the $(f,e_{p,0})$ grid and its dependence on $q_{\rm B}$}
\label{subsec:planetary_response_vs_q}

The representative trajectories shown above provide intuition for individual outcomes, but the response of the system is better assessed across the full grid of initial conditions. Figure~\ref{fig:cbp_maps_q} shows the orbital response in the $(f,e_{p,0})$ plane for a representative unequal-mass binary, $q_{\rm B}=0.30$. This figure is intended as a representative slice through the experiment, rather than as a comparison of static stability domains across mass ratio. The complete set of maps for all explored values of $q_{\rm B}$ is presented in Appendix~\ref{app:maps_full}.

At this mass ratio, survivors occupy a broad region of the near-boundary grid. Unstable outcomes appear preferentially towards smaller initial separations from the stability limit and larger initial planetary eccentricities, corresponding to configurations placed closer to the dynamically fragile part of phase space. Among the surviving planets, the semi-major axis response remains modest over most of the grid: $\Delta a_p/a_{p,0}$ is typically close to zero, although some stable configurations experience inward or outward drifts of a few per cent. The eccentricity response is more structured, with both damping and excitation depending on the initial location in the $(f,e_{p,0})$ plane.

This representative slice illustrates the behaviour seen throughout the survey: the moving stability boundary can change the final separation between the planet and $a_{\rm crit}(t)$ even when the planet itself remains nearly fossilised in semi-major axis. The dependence on binary mass ratio is therefore better quantified statistically, using the full ensemble of surviving systems.

\begin{figure*}
\centering

\begin{minipage}[t]{0.36\textwidth}
    \vspace{0pt}
    \centering
    \includegraphics[width=\linewidth]{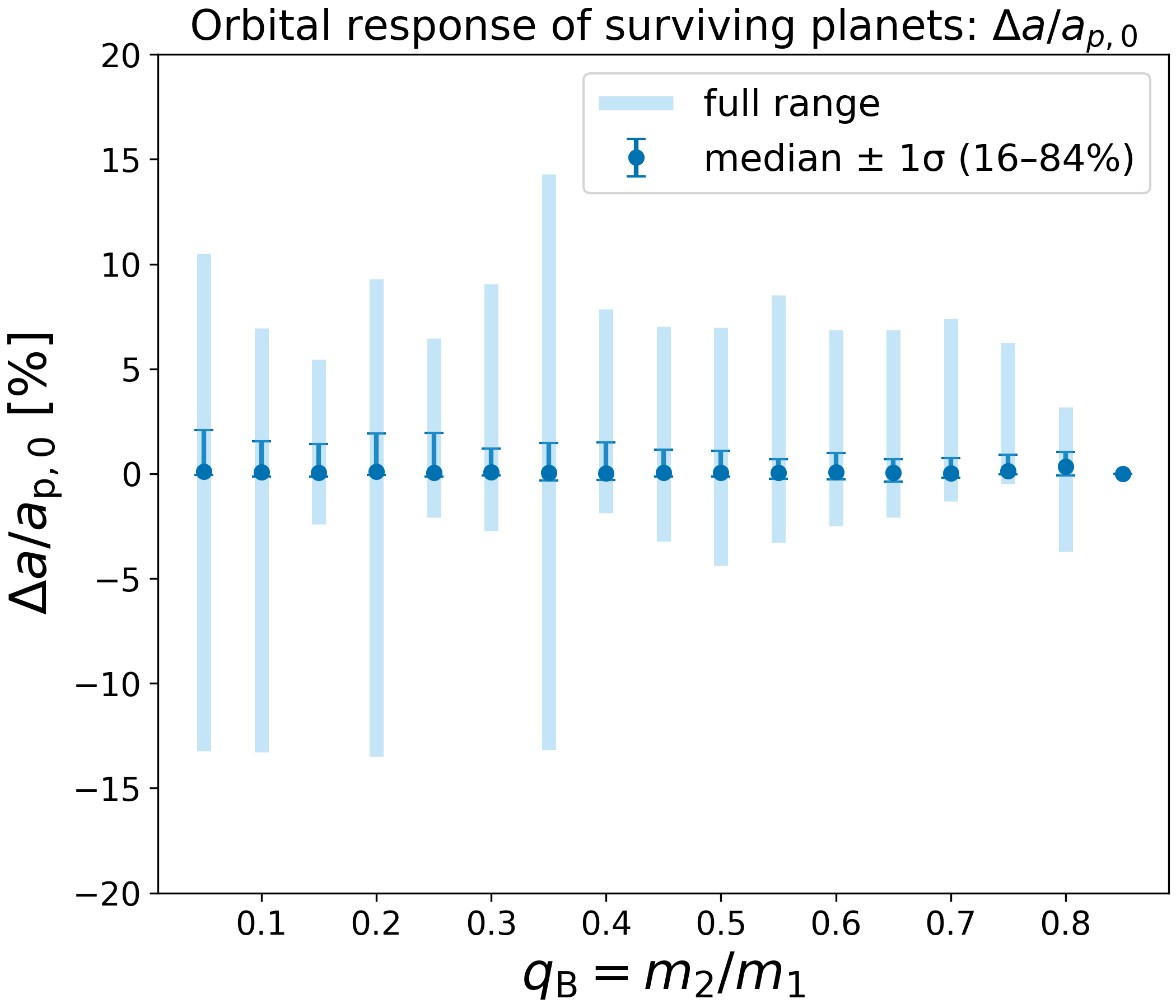}
\end{minipage}
\hfill
\begin{minipage}[t]{0.36\textwidth}
    \vspace{0pt}
    \centering
    \includegraphics[width=\linewidth]{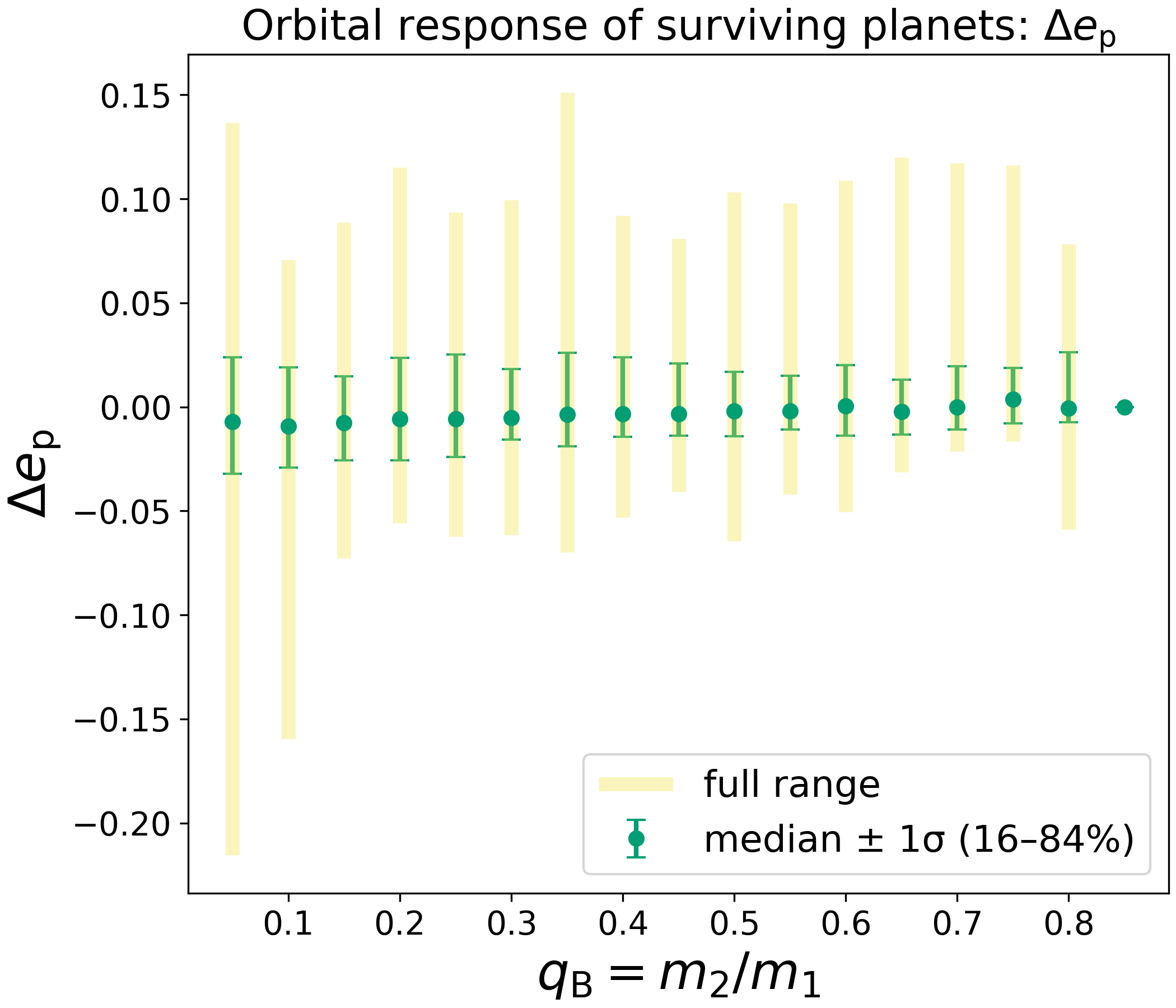}
\end{minipage}
\hfill
\begin{minipage}[t]{0.25\textwidth}
    \vspace{10pt}
    \captionsetup{aboveskip=5pt}
    \captionof{figure}{
    Fractional semi-major axis drift $\Delta a_{p}/a_{p,0}$
    (left) and absolute eccentricity change $\Delta e_{p}$ (right)
    of all surviving planets as a function of the binary mass ratio
    $q_{\rm B}=m_{2}/m_{1}$. The filled markers show the median, and the
    error bars show the 16th--84th percentile range. The translucent vertical
    bars indicate the full min--max span at each $q_{\rm B}$.
}
    \label{fig:Deltas_stats}
\end{minipage}
\end{figure*}

Figure~\ref{fig:Deltas_stats} summarises the response of all surviving planets as a function of $q_{\rm B}$. For each mass-ratio bin, we report the median secular drift of the survivors, together with the 16th--84th percentile range and the full min--max span. The median response remains close to zero in both $a_p$ and $e_p$ over the explored mass-ratio range. This supports the quasi-fossilised picture: for most survivors, the dominant change in the final offset from the stability boundary is produced by the inward motion of $a_{\rm crit}$ rather than by substantial planetary migration.

The percentile ranges and full spans nevertheless show that the response is not identical for all survivors. A minority of stable planets undergoes larger, but non-destructive, changes in semi-major axis or eccentricity. These cases correspond to trajectories that remain bound while sampling more dynamically active regions during binary contraction and eccentricity damping. Thus, the median response captures the typical quasi-fossilised behaviour, whereas the full range highlights the diversity of stable outcomes within the same evolutionary experiment.

\subsection{Planetary fates under binary contraction}
\label{subsec:fates_vs_q}

The previous subsection focused on the orbital response of the surviving systems. We now ask how binary contraction redistributes the initial circumbinary population among the possible dynamical outcomes: survival, ejection, or Roche-limit crossing. We also examine whether the losses occur before, during, or after the imposed evolution of the binary. For each $q_{\rm B}$, we classified all integrations into three physical outcomes: survival (bound at the end of the integration), ejection (unbound escape), and collision (with either stellar component). The `error' category groups integrations terminated by numerical or bookkeeping failures and is negligible in practice.

\begin{figure*}
  \centering

  \includegraphics[width=0.32\textwidth]{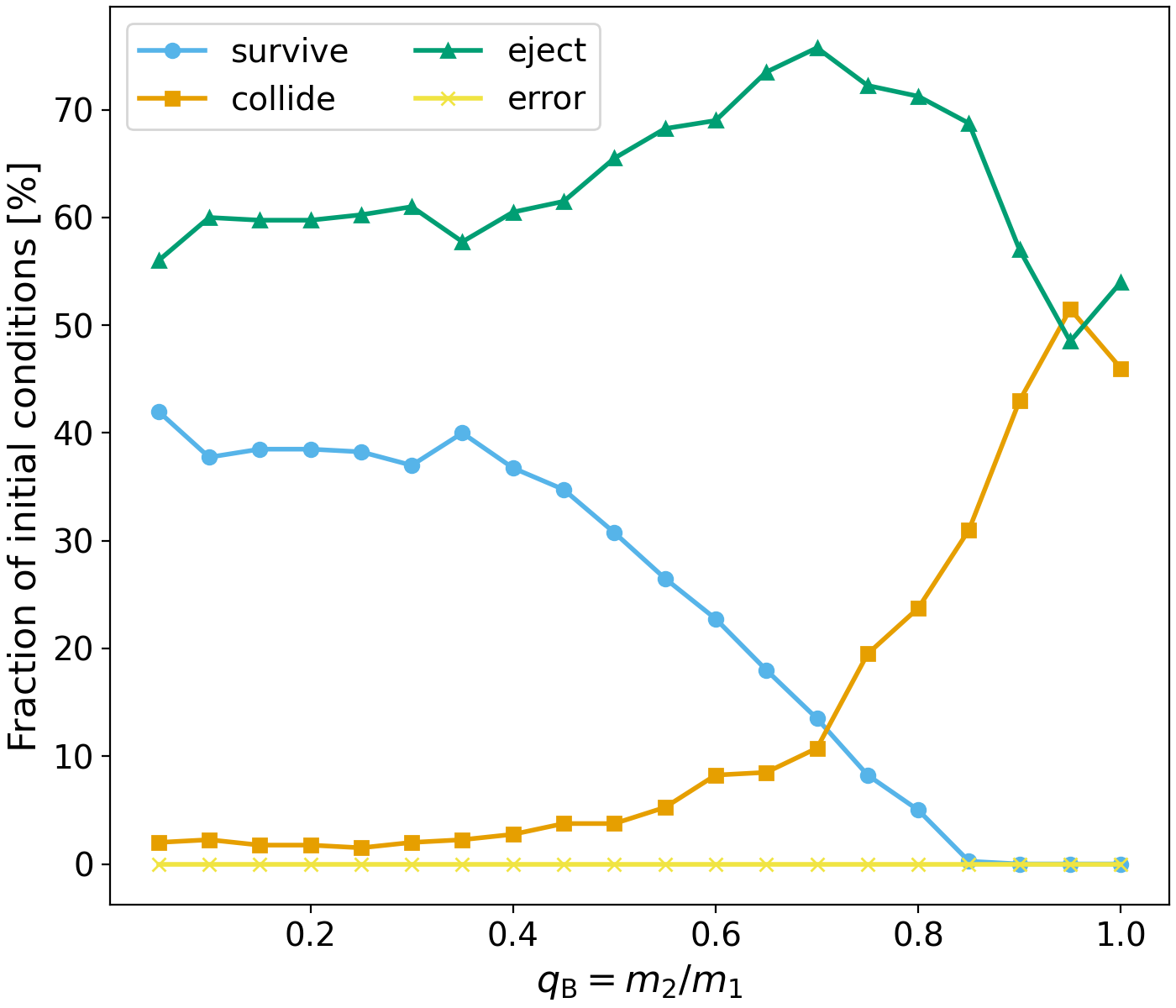}
  \hfill
  \includegraphics[width=0.32\textwidth]{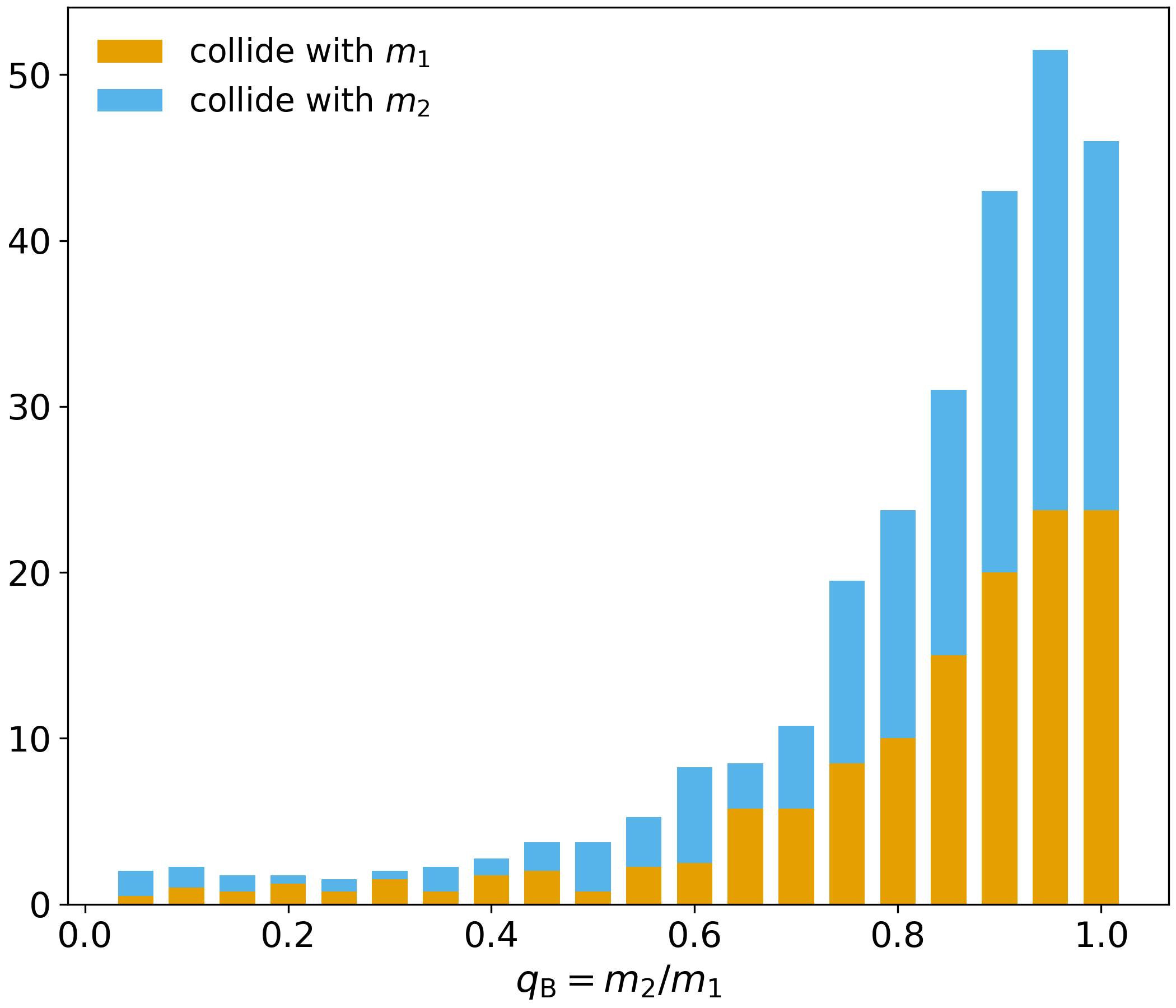}
  \hfill
  \includegraphics[width=0.32\textwidth]{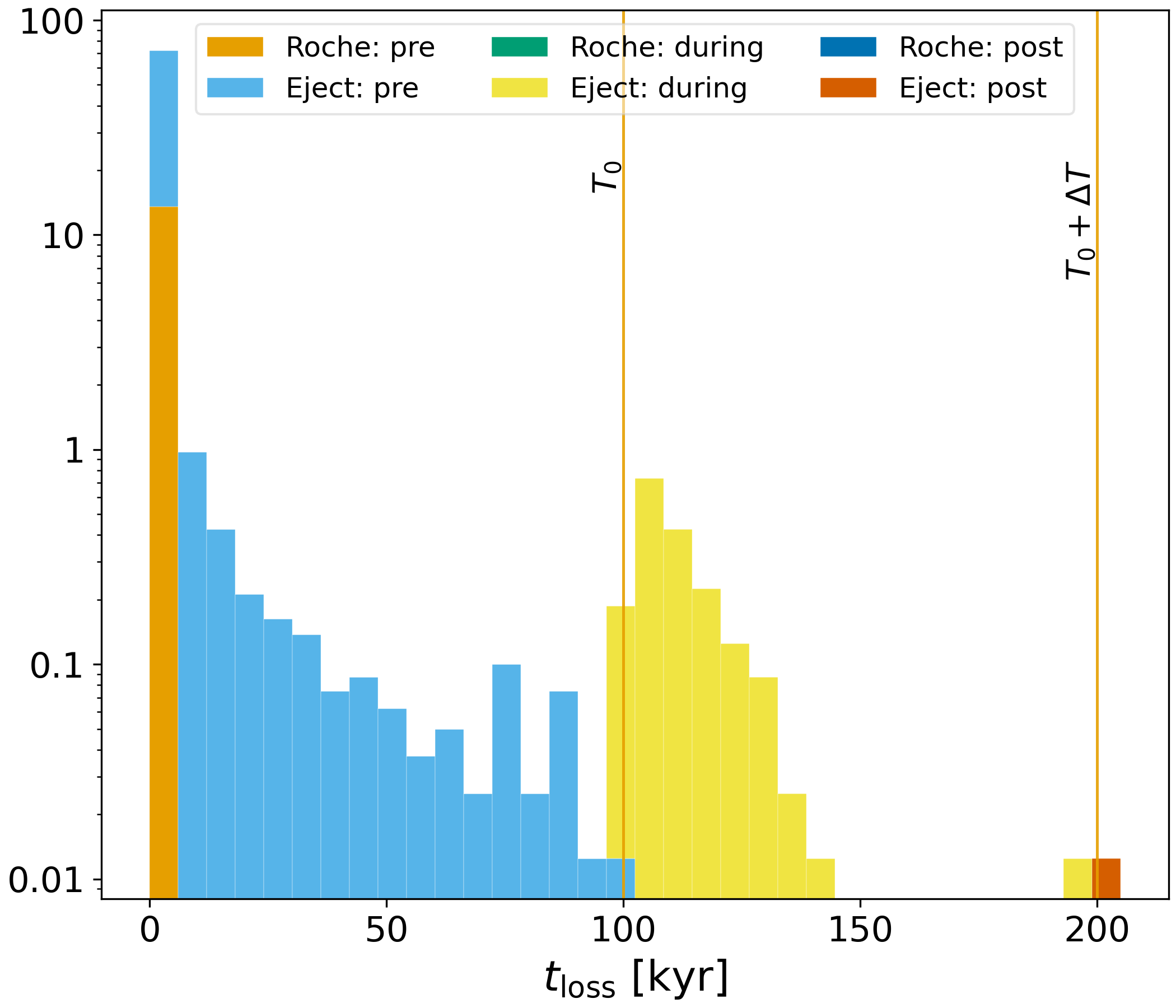}

  \caption{
   {Fate and loss channels of CBPs in the tidal-evolution runs.
    {Left:} Outcome fractions as a function of the binary mass ratio $q_{\rm B}=m_2/m_1$.
    {Centre:} Roche losses split by impacts with the primary ($m_1$) and secondary ($m_2$).
    {Right:} Fraction of initial conditions per time bin, split by loss channel and evolutionary phase. The vertical lines mark $T_0$ and $T_0+\Delta T$.}
  }
  \label{fig:fates_q}
\end{figure*}

Figure~\ref{fig:fates_q} shows the final fate budget as a function of $q_{\rm B}$. The survival fraction decreases overall with increasing mass ratio, while the loss budget is dominated by ejections over a broad range of $q_{\rm B}$. For the largest mass ratios, Roche collisions also become increasingly frequent. Thus, binaries closer to equal mass are more efficient at dynamically filtering the initially sampled circumbinary region, although the relative importance of ejections and collisions depends on the part of parameter space considered.

The outcome statistics separate into two broad regimes. At low-to-intermediate mass ratios ($q_{\rm B}\lesssim0.5$), survival remains common and losses are mainly through ejections, with Roche collisions still rare. At higher mass ratios ($q_{\rm B}\gtrsim0.6$), survival drops sharply and the loss budget becomes more destructive, with ejections remaining important and Roche collisions increasing towards the near-equal-mass regime. The near-equal-mass limit is therefore the most strongly filtered part of the explored grid, where most of the initially sampled near-boundary population is removed.

The collision channel also depends on which stellar component is impacted. Impacts occur onto both stars, but the relative contribution of each component varies with $q_{\rm B}$, and the total collision fraction increases strongly as $q_{\rm B}\rightarrow 1$. This raises the possibility that contraction-driven instabilities could leave differential chemical `pollution' signatures between the two stellar components if one of them preferentially accretes planetary material. We return to this point in Sect.~\ref{sec:discussion}.

The final fate budget alone does not indicate whether the lost planets were already short-lived before the imposed binary evolution, or whether they were destabilised during or after the contraction phase.  {To separate these contributions, we recorded the time of loss relative to the forcing interval. The right panel of Fig.~\ref{fig:fates_q} shows the distribution of loss times, split by channel and evolutionary phase:} before the onset of forcing ($t<T_0$), during binary contraction and eccentricity damping ($T_0 \leq t \leq T_0+\Delta T$), and after the forcing has ended ($t>T_0+\Delta T$).

Because the initial grid deliberately samples the near-boundary region, including configurations close to or slightly inside the initial stability limit, a large fraction of the total losses is expected to occur early and to reflect the classical static marginal-stability structure. To isolate the time-dependent contribution, we compare each evolving-binary run with its matched fixed-binary control in Appendix~\ref{app:excess_loss}. This identifies the transition $S_{\rm ctrl}\rightarrow L_{\rm tides}$, namely planets that survive around the fixed binary but are lost when binary contraction and eccentricity damping are imposed. This matched-control comparison shows that the excess-loss
component is modest, reaching a few per cent of the valid initial conditions, and that all induced losses are ejections occurring during the forcing interval (Appendix~\ref{app:excess_loss}).

A large fraction of the losses occurs before $T_0$. This shows that part of the near-boundary grid samples configurations that are intrinsically short-lived, even before substantial binary evolution takes place. We interpret this component as the static or marginal-stability contribution to the loss budget.

A second group of losses occurs during the forcing interval, and is dominated by ejections. These events are the clearest signature that the moving stability boundary contributes to planet removal during binary contraction and eccentricity damping. In this sense, the fate budget is not only a map of the underlying static instability structure. Part of it is linked in time to the imposed evolution of the binary. Losses after $T_0+\Delta T$ are rare in our integrations, suggesting that most of the filtering occurs either before the forcing begins or during the contraction phase itself. The panels in Fig.~\ref{fig:fates_q} therefore provide complementary diagnostics. The left and centre panels quantify how the final survival, ejection, and collision fractions depend on $q_{\rm B}$, while the right panel separates the losses associated with the initial near-boundary stability structure from those temporally linked to binary evolution.

\subsection{Comparison with the observed circumbinary planet population}
\label{subsec:obs_comparison}

\begin{figure*}
  \centering
  \includegraphics[width=0.96\textwidth]{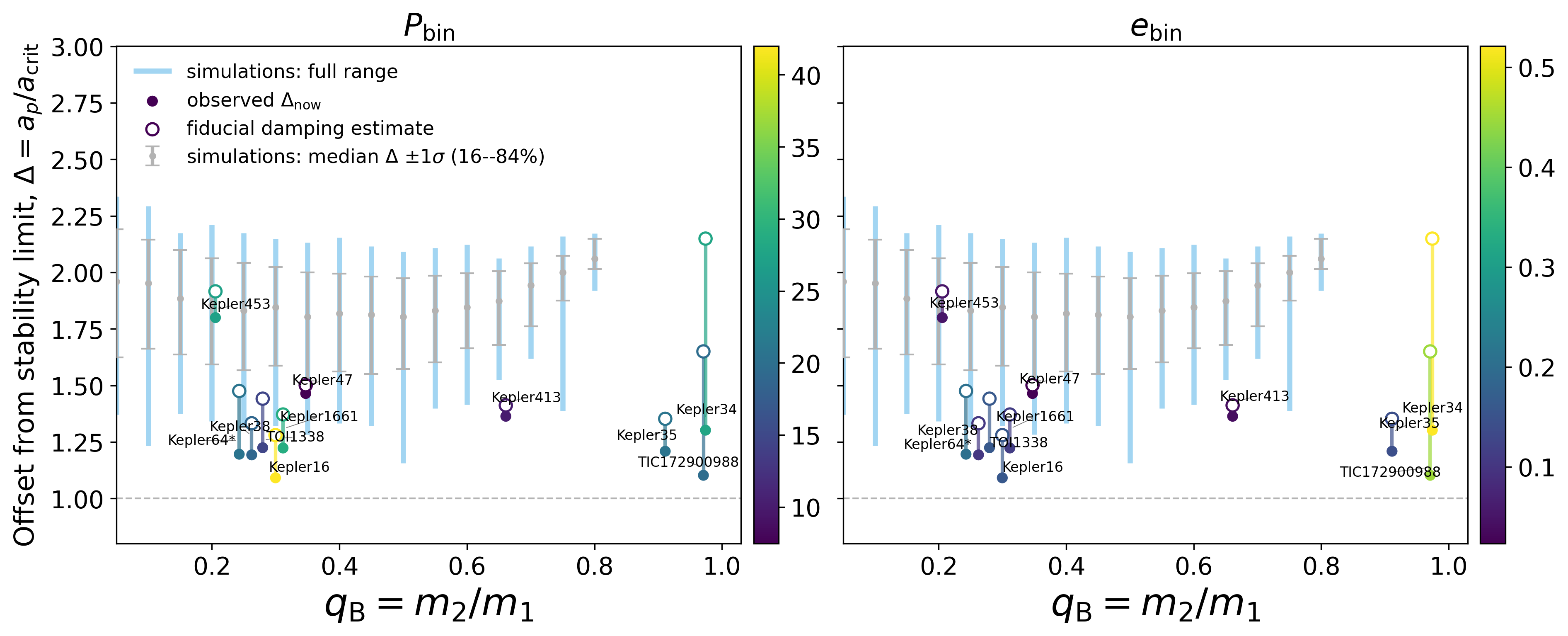}
  \caption{
  Comparison between synthetic survivors and observed CBPs. The black symbols show the median final offset $\Delta_f=a_{p,\mathrm{fin}}/a_{{\rm crit},f}$ of all valid synthetic survivors at each $q_{\rm B}$, with error bars indicating the 16th--84th percentile range. The vertical light-blue bars show the full range of $\Delta_f$. The filled circles show the observed present-day offsets, $\Delta_{\rm now}$, excluding BEBOP-3 and Kepler-1647. The open circles show the corresponding fiducial damping estimates, computed following the prescription described in Sect.~\ref{sec:bin_evol}. The colours indicate $P_{\rm bin}$ and $e_{\rm bin}$ in the left and right panels, respectively.
}

  \label{fig:populations}
\end{figure*}

We now compare our synthetic architectures with the observed CBPs. Figure~\ref{fig:populations} shows the final offset of the surviving synthetic population, $\Delta_f \equiv a_{p,\mathrm{fin}}/a_{{\rm crit},f}$, as a function of the binary mass ratio. For each $q_{\rm B}$ bin, black symbols show the median value of $\Delta_f$ among all valid survivors, with error bars spanning the 16th--84th percentile range. The vertical light-blue bars indicate the full range of $\Delta_f$ reached by the same surviving systems. The synthetic survivor distribution is displaced towards $\Delta_f>1$ over the explored mass-ratio range. Considering all valid synthetic survivors, we obtain a global median final offset of $\Delta_f = 1.86^{+0.18}_{-0.26}$, where the uncertainties correspond to the 16th--84th percentile range. This displacement follows from the inward retreat of $a_{\rm crit}(t)$ during binary contraction and partial circularisation. Since the typical surviving planet undergoes only modest changes in semi-major axis (Fig.~\ref{fig:Deltas_stats}), the increase in $\Delta$ is produced mainly by the motion of the stability boundary rather than by large-scale planetary migration.

For the observed systems, the filled circles show the present-day offset,
$\Delta_{\rm now}=a_p/a_{\rm crit}$.
We exclude the extreme systems BEBOP-3 and Kepler-1647 from this comparison to focus on the main compact transiting and radial-velocity population. For each observed planet, we also computed a simple fiducial damping estimate. In this estimate, the planetary semi-major axis was kept fixed, while the binary eccentricity was damped by the same factor used in the fiducial simulations, $\exp(-\Delta T/\tau_e)$. The binary semi-major axis was then updated by conserving $a_{\rm bin}(1-e_{\rm bin}^2)$, after which we recomputed $a_{\rm crit}$ and the corresponding offset,
$\Delta_{\rm damp}$. These estimates are shown as open circles connected to the observed values by vertical segments.

Applying this prescription to the observed sample shifts the median offset from
$\Delta_{\rm now}=1.22^{+0.18}_{-0.07}$
to
$\Delta_{\rm damp}=1.44^{+0.32}_{-0.10}$.
The median displacement,
$\Delta_{\rm damp}-\Delta_{\rm now}=0.15^{+0.24}_{-0.07}$,
shows that the observed systems move in the expected direction under boundary retreat alone. This estimate should not be interpreted as a reconstruction of the future evolution of each system, but as a fossilised-orbit diagnostic of the offset produced by damping the binary eccentricity while keeping the planetary orbit fixed. The effect is not uniform: eccentric and near-equal-mass binaries, such as Kepler-34 and TIC~172900988, show the largest displacement, whereas weakly eccentric systems such as Kepler-47 or Kepler-413 move only slightly. A full-circularisation estimate, in which $e_{\rm bin}$ is set to zero, gives a larger median value, $\Delta_{\rm circ}=1.52^{+0.48}_{-0.09}$, and should be interpreted as an upper-limit diagnostic rather than as the fiducial comparison.

The observed planets, even after this fiducial damping estimate, mostly lie below the median synthetic survivor distribution in this end-member experiment. This does not contradict the boundary-retreat picture. It suggests that real systems may have experienced smaller degrees of binary evolution, different initial architectures, disc-driven migration and damping, or later dynamical evolution. In this sense, binary boundary retreat provides a migration-free contribution to the observed offsets, but it need not be the only process shaping the present-day population.

The colour coding by present-day $P_{\rm bin}$ and $e_{\rm bin}$ provides a qualitative check on possible tidal histories. Short-period or eccentric binaries are natural systems in which boundary retreat could produce a larger change in $\Delta$, although the current sample is too small and covariant in $(q_{\rm B},P_{\rm bin},e_{\rm bin})$ to draw a firm trend. A more direct comparison requires coupling the boundary-retreat mechanism explored here with formation and migration models, as well as with observational selection effects.

\section{Discussion}
\label{sec:discussion}

In this work, we studied the effect of binary contraction and eccentricity damping on CBPs initially located near the dynamical stability boundary. Our results show that many surviving planets undergo only modest changes in semi-major axis while the critical stability limit retreats inward. The resulting increase in $\Delta=a_p/a_{\rm crit}$ therefore reflects mainly the motion of the stability boundary, rather than large-scale planetary migration. This provides a simple pathway by which CBPs can be displaced from marginal stability after their formation and early disc-driven evolution.

Our work complements recent studies proposing additional channels that may operate concurrently. \citet{Farhat2025} outline a secular pathway in which tides and general relativity modify the binary apsidal precession, potentially sweeping a secular resonance through the circumbinary region. This process can excite planetary eccentricities while leaving more distant survivors with a characteristic apsidal structure. In parallel, \citet{Mogan2025} show that binary orbital decay can strongly reduce the probability of detecting transiting CBPs. Part of the apparent desert around tight binaries may therefore reflect geometric detectability, even when planets survive at larger separations.

However, we did not attempt to identify or track the secular resonances described by \citet{Farhat2025}. Our experiment instead targeted the near-boundary circumbinary region, where planets are expected to be most sensitive to the inward motion of $a_{\rm crit}(t)$. This region is especially relevant because it defines the observed inner gap and its width. In compact, nearly circular binaries, the present-day lack of marginally stable planets may partly reflect a past retreat of the stability boundary, with the final offsets depending on the binary properties, including $q_{\rm B}$. Since these offsets also affect the probability of observing transiting CBPs, this links our boundary-retreat picture with the detectability scenario discussed by \citet{Mogan2025}. In this sense, our study is complementary to both resonance-based and detectability-based explanations. Rather than following a specific secular pathway or modelling transit selection, we quantified how a moving stability boundary can filter, displace, or remove planets initially located close to marginal stability.

A consistent interpretation is that the CBP desert is unlikely to have a single origin. It may instead reflect a combination of physical sculpting, including instability, ejection, and collision channels, together with geometric selection effects associated with transiting configurations. In this multi-channel picture, our contribution is to isolate a simple internal mechanism: the inward motion of $a_{\rm crit}(t)$ during binary contraction and eccentricity damping.

\subsection{Binary contraction and the moving circumbinary stability boundary}
\label{subsec:binary_contraction_boundary}

The surviving planets support the quasi-fossilised picture. Across the explored mass-ratio range, median changes in semi-major axis and eccentricity remain small (Fig.~\ref{fig:Deltas_stats}), so the final increase in $\Delta$ is driven mainly by the retreat of $a_{\rm crit}(t)$.

The balance between survival and loss depends on the binary mass ratio. This is consistent with previous studies showing that the extent and structure of stable circumbinary regions depend on the binary parameters, in particular $q_{\rm B}$ and $e_{\rm bin}$ \citep[e.g.][]{Dvorak1986,Dvorak1989,Holman_Wiegert1999,Pilat-Lohinger2003,Quintana2006,Morais2012}. In our experiment, however, the relevant point is not only the static location of the stability boundary but how the near-boundary population responds when that boundary moves during binary contraction and eccentricity damping. At low and intermediate $q_{\rm B}$, survival remains common. Towards near-equal-mass binaries, the experiments show
stronger dynamical filtering, with ejections and Roche
collisions becoming increasingly frequent as
$q_{\rm B}\to 1$  (Fig.~\ref{fig:fates_q}). The surviving population in this regime is therefore a more strongly selected subset of the initial near-boundary population.

The timing of the losses further separates the static and time-dependent components of this filtering process.  {The right panel of} Fig.~\ref{fig:fates_q} shows that a large fraction of losses occurs before the forcing phase begins. These losses correspond to configurations that are already short-lived in the initial near-boundary grid, and should be interpreted as the static or marginal-stability contribution to the loss budget. A second component occurs during the forcing interval, dominated by ejections. These events are temporally associated with the imposed binary evolution, when the retreat of $a_{\rm crit}(t)$ and the changing binary potential modify the surrounding phase space. Losses after the forcing phase are comparatively rare, suggesting that most of the filtering occurs either before substantial binary evolution begins or during binary contraction and eccentricity damping.

Within this controlled framework, CBPs may preserve a macroscopic record of binary evolution. The relevant signatures are the increase in $\Delta$ among quasi-fossilised survivors, the mass-ratio-dependent survival and loss budget, and the timing of losses relative to the imposed binary contraction and eccentricity-damping phase. These signatures do not provide a direct measurement of tidal dissipation efficiencies or timescales, but they show that binary evolution can leave an observable imprint on the distribution of planetary orbits relative to $a_{\rm crit}$.

The Jupiter-mass test in Appendix~\ref{app:planet_mass} shows that these survival statistics are robust to the adopted planet mass, although the partition between ejections and Roche losses shows some sensitivity for nearly equal-mass binaries. These trends are also only weakly sensitive to the adopted forcing timescale: over the factor-of-ten range tested, faster forcing leaves slightly more survivors and somewhat smaller final offsets, without changing the population-level trends (Appendix~\ref{app:timescale}).

\subsection{A late-time ejection channel for circumbinary planets}

The sharp decline in survival towards high $q_{\rm B}$, together with the persistence of ejections and the rise of Roche collisions, suggests that binary contraction and eccentricity damping can provide a late-time pathway for CBP loss. Unlike early disc-driven scattering or primordial dynamical instabilities, this channel can operate after disc dispersal and does not require external perturbers. Planet removal arises from the gradual reshaping of the stable circumbinary region as $a_{\rm crit}(t)$ moves inward.

This result is relevant because close binaries tend to favour higher mass ratios (Sect.~\ref{sec:introduction}). If primordial CBPs were common around such systems, a fraction of them may have been exposed to the enhanced
instability regime found here. The currently known transiting CBPs may therefore be biased towards a selected survivor population, rather than an unbiased sample of the underlying circumbinary population.

In this framework, the surviving and ejected populations are linked by the same internal process. The tidal prescription controls when and how efficiently the binary evolves, but the qualitative pathway is set by the retreat of
$a_{\rm crit}(t)$. This makes contraction-driven ejection complementary to early channels of free-floating planet production. For example, disc-mediated scattering may preferentially eject planets in low-$q_{\rm B}$ systems \citep{Calovic2026}, whereas the mechanism explored here becomes more effective, in our simulations, towards high $q_{\rm B}$. Embedded disc-fragmentation scenarios can also eject planets during the early stages of binary and planet formation, contributing to the microlensing free-floating-planet population and possibly to a bottom-heavy mass function \citep[e.g.][]{Nayakshin2026,Zhang2026}.  {Multi-planet architectures may further modify this picture. \cite{Grane2026} show that planet–planet interactions can either maintain stable resonant configurations or enhance instability, depending on the binary parameters and migration history. At later stages, tide-driven secular pathways may amplify local eccentricity excitation into system-wide instability through planet–planet scattering, leading to collisions and ejections \citep{Liu2026}.}

Our simulations also quantify the redistribution of an initially stable circumbinary population among survival, ejection, and Roche-loss channels. However, we did not model the subsequent physical evolution of the Roche-loss channel, such as partial disruption, debris formation, re-accretion, or observable abundance anomalies \citep{Sucerquia2025,Montesinos2026}. Such processes could connect planet loss to chemical pollution of one or both stellar components \citep[e.g.][]{Oh2018}. High-precision differential abundance studies of binary stars show that this type of signature can be detectable: \citet{Rathsam2026} found refractory-element, Li, and Be differences in the HD~129171/HD~129209 pair consistent with the engulfment of rocky material by one component. Although that system is not a circumbinary-planet analogue, it illustrates that planet ingestion can leave measurable chemical imprints in binary pairs. Precise abundance measurements in double-lined spectroscopic binaries remain observationally challenging, but they could provide an independent test of the Roche-loss channel explored here. Recent time-domain detections of catastrophic star--planet interactions, such as ZTF~SLRN-2020 \citep{Lau2025}, show that this broader class of events is becoming observationally accessible.

\subsection{Theoretical expectations and observational tests}
\label{subsec:observational_tests}

The boundary-retreat picture leads to a simple observational expectation: the innermost surviving CBPs should carry information about both their formation and migration history
and the later evolution of the host binary. In systems that experienced stronger contraction and eccentricity damping, the stability boundary is expected to have moved farther inward, increasing the offset $\Delta=a_p/a_{\rm crit}$ for surviving near-boundary planets. At the same time, the same process can remove part of the initially marginal population through ejections or collisions. The relevant test is therefore the joint distribution of $(q_{\rm B},P_{\rm bin},e_{\rm bin},\Delta)$ rather than $\Delta$ alone.

Our fiducial damping estimate for the observed compact transiting and radial-velocity CBPs illustrates this diagnostic. Keeping the planetary semi-major axes fixed and applying the same eccentricity-damping strength as in the simulations shifts the median offset from
$\Delta_{\rm now}=1.22^{+0.18}_{-0.07}$
to
$\Delta_{\rm damp}=1.44^{+0.32}_{-0.10}$.
This does not reconstruct the history of individual systems, but it shows how the observed architecture moves in the expected direction under boundary retreat alone. Since larger offsets imply longer planetary periods and lower geometric transit probabilities, this dynamical displacement is naturally coupled to detectability-based explanations of the CBP desert. This also affects how target samples are interpreted: less tidally evolved binaries may retain planets closer to marginal stability, whereas strongly evolved compact binaries may host survivors at longer periods or may have already lost part of their initial near-boundary population.

Planetary eccentricity may provide an additional diagnostic. Close CBPs sample a non-Keplerian, time-dependent potential near $a_{\rm crit}$, where forced eccentricities and weakly stable regions can remain important. In our maps, some near-boundary survivors are damped whereas others are excited during binary contraction and eccentricity damping. Thus, $e_p$, together with $\Delta$, may help distinguish quasi-fossilised architectures from dynamically filtered ones as the observed CBP sample grows. However, in hierarchical triple systems, an outer stellar companion may further modify this picture by altering the secular forcing of the CBP and, in some configurations, producing regions of reduced forced eccentricity \citep{Gianuzzi2026}.

\subsection{Relation to previous tidal models}

Previous work has considered the role of tides in CBP systems from different perspectives. One possibility is direct star--planet tidal coupling, in which the binary raises tides on the CBP. The detailed analysis of \citet{Zoppetti2020} showed that this effect is dynamically negligible for typical CBPs, producing changes in planetary semi-major axis and eccentricity below $10^{-4}$ over gigayear timescales. This implies that direct star--planet tides are unlikely, by themselves, to shift the inner circumbinary cavity or explain the observed offsets between $a_p$ and $a_{\rm crit}$.

Our mechanism is different. We do not rely on direct tidal evolution of the planet. Instead, tides act on the central binary, whose contraction and eccentricity damping move the stability boundary $a_{\rm crit}(t)$. The planet then responds to the evolving binary potential and to the retreat of the stable phase space. This explains why, in our simulations, many surviving planets show only modest changes in $a_p$, while their offset $\Delta=a_p/a_{\rm crit}$ increases substantially.

A second complementary pathway involves secular resonances associated with the tidal evolution of the binary. 
\citet{Farhat2025} show that tides and relativistic precession can modify the apsidal evolution of compact binaries, potentially sweeping secular resonances through the circumbinary region and exciting planetary eccentricities. 
In the compact multi-planet systems studied by \citet{Liu2026}, tidal decay of an eccentric inner binary can further couple the apsidal precession of the binary and the planets, leading to resonance advection, eccentricity growth, orbit crossing, collisions, and ejections. 
These mechanisms differ from ours in two important ways: they rely on resonant secular coupling, and the most unstable outcomes are enhanced in multi-planet architectures. 
Our simulations include only one planet per system and do not attempt to track apsidal resonance capture. They therefore isolate the boundary-retreat channel rather than the resonance-driven instability channel.

For the fiducial systems explored here, however, this apsidal coupling is unlikely to operate. At $a_{\rm bin}=0.3$ AU, the relativistic precession rate is of order $\dot{\varpi}_{\rm GR}\sim 5\times10^{-6}\ {\rm rad\,yr^{-1}}$, whereas the binary-driven apsidal precession of planets near the stability boundary is typically $\dot{\varpi}_p\sim10^{-2}$--$10^{-1}\ {\rm rad\,yr^{-1}}$. The two rates therefore remain separated by several orders of magnitude throughout our imposed evolution. Moreover, approximate conservation of $a_{\rm bin}(1-e_{\rm bin}^2)$ limits the binary contraction from $0.30$ AU to $0.252$ AU even for complete circularisation. The apsidal-resonance channel is therefore expected to become relevant only for substantially more compact evolutionary tracks, such as those starting from much larger binary eccentricities or involving additional angular-momentum loss.

These complementary mechanisms suggest that tidal evolution can affect circumbinary architectures through at least two indirect routes: by moving the stability boundary, as explored here, and by sweeping or locking secular resonances, as in \citet{Farhat2025} and \citet{Liu2026}. Both are consistent with the conclusion of \citet{Zoppetti2020} that direct star--planet tides are too weak to dominate the evolution of CBPs. Recent frameworks that couple stellar evolution to the frequency-dependent tidal response provide a route towards modelling the binary orbital evolution self-consistently as the stellar structure changes \citep{Narayan2026}. Future models combining realistic binary tides, relativistic precession, and multi-planet dynamics will be needed to determine which pathway dominates for different binary and planetary architectures.

\section{Conclusions}
\label{sec:conclusions}

We explored whether the observed offset of CBPs from marginal stability,
$\Delta=a_p/a_{\rm crit}>1$, can partly record the later evolution of the central binary. Our controlled experiments show that circumbinary architectures can retain this imprint even when the planets themselves migrate very little. In many surviving systems, the relevant change is not a large displacement of $a_p$, but the inward motion of $a_{\rm crit}(t)$ during prescribed binary contraction and eccentricity damping. Considering all valid synthetic survivors, this produces a global median final offset of
$\Delta_f=1.86^{+0.18}_{-0.26}$.

The same mechanism also acts as a dynamical filter. At low-to-intermediate $q_{\rm B}$, survival remains common and many planets remain quasi-fossilised. Towards near-equal-mass binaries, the circumbinary region becomes more fragile: the survival fraction drops, ejections increase, and Roche collisions become non-negligible. The timing of the losses shows that part of the loss budget is associated with initially marginal configurations, while another part occurs during the imposed binary-evolution phase.

The comparison with the observed circumbinary population should be understood as illustrative rather than as a direct reconstruction of individual systems. Applying our fiducial damping prescription to the observed compact transiting and radial-velocity sample, while keeping the planetary semi-major axes fixed, shifts the median offset from
$\Delta_{\rm now}=1.22^{+0.18}_{-0.07}$
to
$\Delta_{\rm damp}=1.44^{+0.32}_{-0.10}$.
This shows that the observed systems move in the expected direction under boundary retreat alone, although real architectures probably combine different degrees of binary evolution, different initial conditions, disc-driven migration and damping, and observational selection effects.

Our results identify binary tidal evolution as a plausible internal sculpting mechanism for circumbinary architectures. Within this phenomenological Newtonian framework, a moving stability boundary, a mass-ratio-dependent survival and loss budget, and a diagnostic displacement of the observed offsets emerge as possible signatures of binary evolution. More realistic tidal prescriptions, relativistic precession, multi-planet dynamics, and detectability modelling would allow us to better connect this mechanism with population-level predictions in the context of ongoing and future CBP surveys.

\begin{acknowledgements}
   This project was supported by the European Research Council (ERC) under the European Union Horizon Europe research and innovation programme (grant agreement No. 101042275, project Stellar-MADE). We also thank MSP and the Stellar-MADE team for their kind support.
\end{acknowledgements}


\bibliographystyle{aa}
\vspace{-0.4cm}
\bibliography{references.bib}

\begin{appendix}
\onecolumn
\nolinenumbers

\section{Orbital response across the $(f,e_{p,0})$ grid for different binary mass ratios}
\label{app:maps_full}

\begin{figure}[H]
  \centering
  \includegraphics[width=\textwidth]{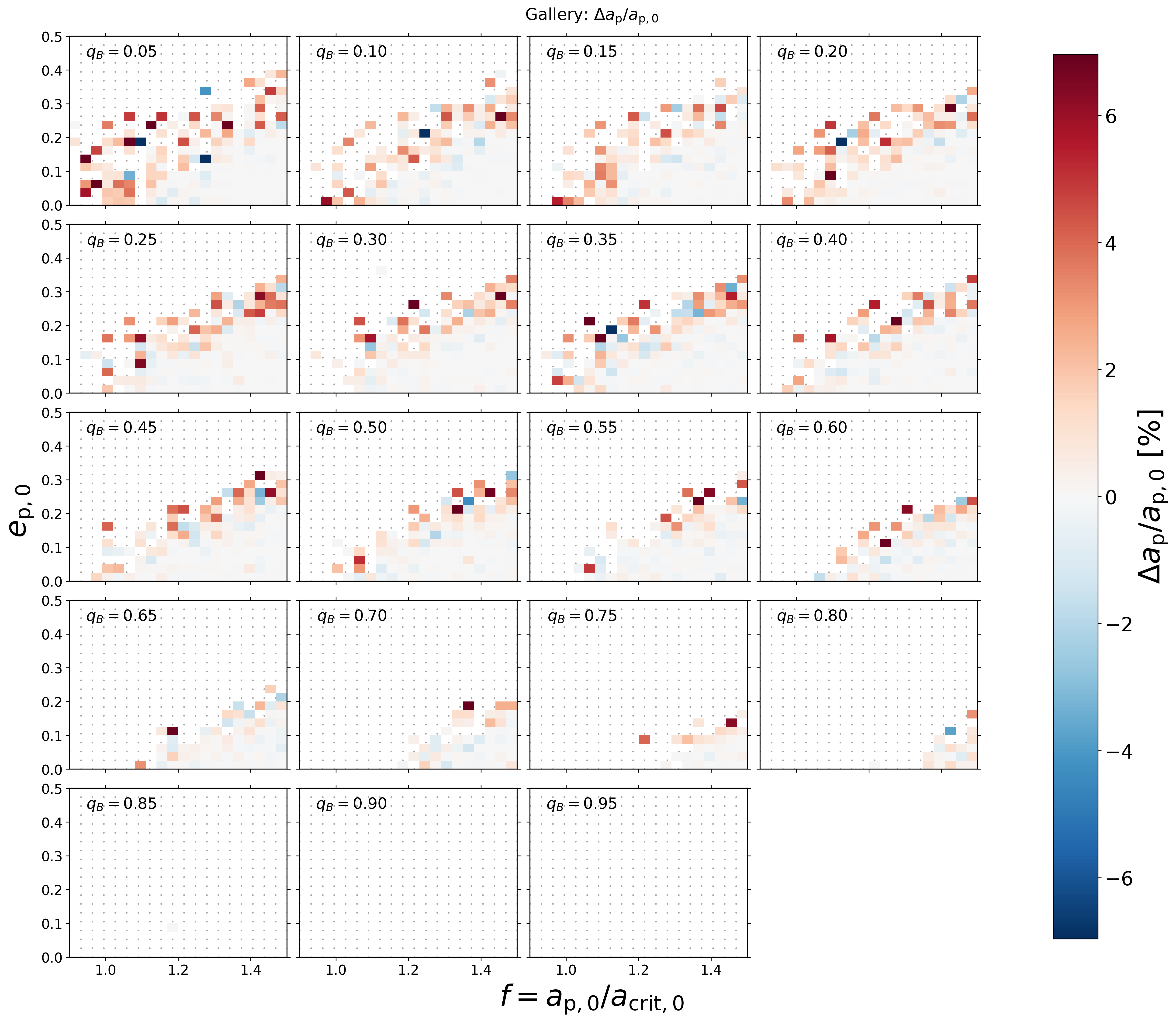}
  \caption{
  Gallery of semi-major axis response maps for surviving CBPs across the $(f,e_{p,0})$ grid, where $f \equiv a_{p,0}/a_{\rm crit,0}$, for a fixed initial binary configuration ($a_{\rm bin,0}=0.30$~AU, $e_{\rm bin,0}=0.40$) and planet mass $m_p=3\times10^{-6}\,M_\odot$.
  Each tile corresponds to a different binary mass ratio $q_{\rm B}=m_2/m_1$, indicated inside the panel, and shows the fractional semi-major axis change $\Delta a_p/a_{p,0}$ (in percent) measured between the beginning and the end of the integration.
  Only surviving systems are colour-coded; unstable outcomes, including ejections and Roche-limit crossings, are marked by grey dots.
  The colour scale is unified across all tiles to enable direct visual comparison as a function of $q_{\rm B}$.
  }
  \label{fig:cbp_maps_q_full_da}
\end{figure}

\begin{figure}[H]
  \centering
  \includegraphics[width=\textwidth]{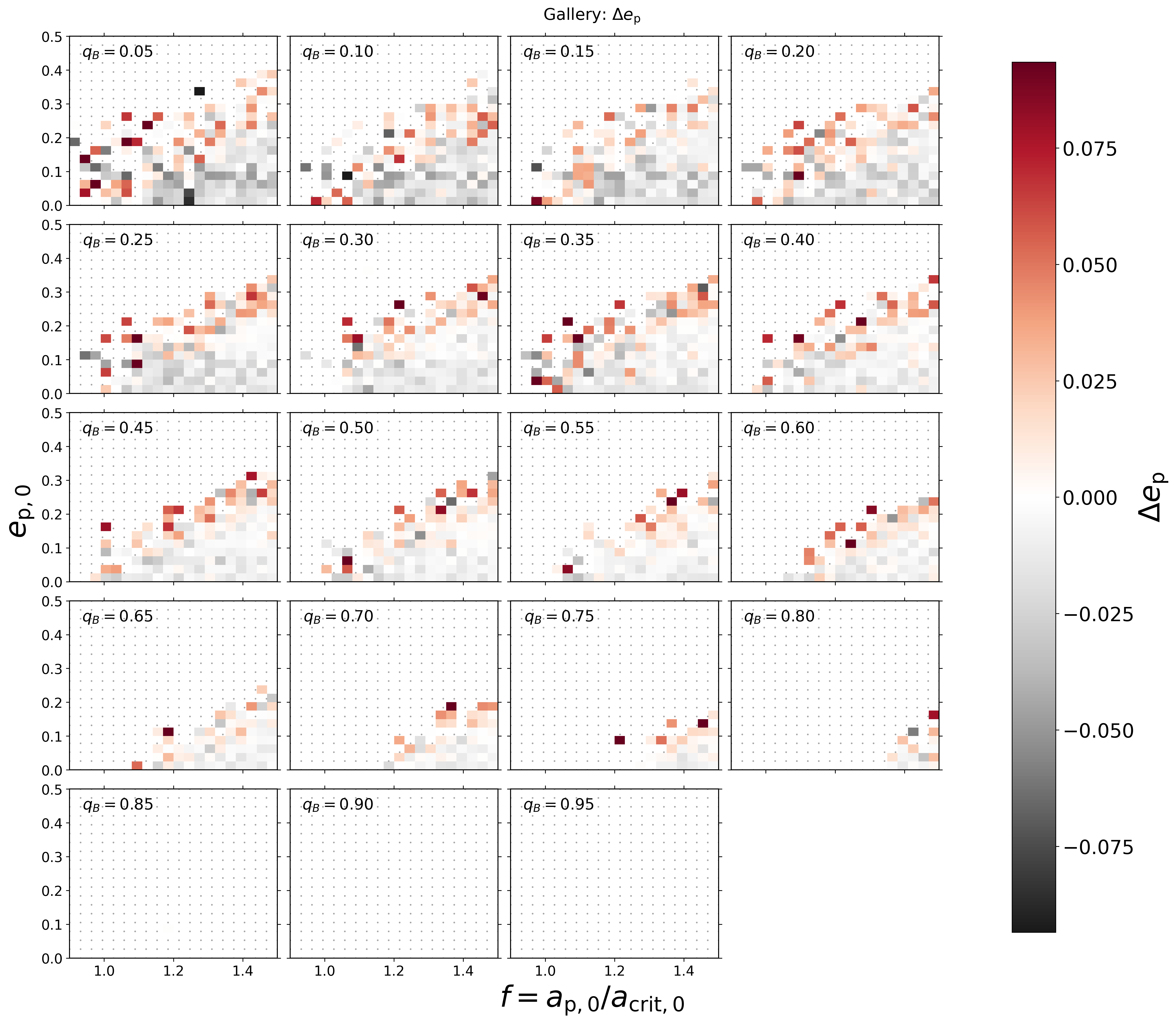}
  \caption{
  Same as Fig.~\ref{fig:cbp_maps_q_full_da} but for the eccentricity response $\Delta e_p$.
  The colour scale is again unified across all tiles, enabling a direct comparison of the eccentricity evolution across binary mass ratios.
  }
  \label{fig:cbp_maps_q_full_de}
\end{figure}

To complement the representative maps shown in Sect.~4.2, in this Appendix we present the full set of orbital-response maps for our fiducial binary configuration ($a_{\rm bin,0}=0.30$~AU, $e_{\rm bin,0}=0.40$) across the explored binary mass ratios.
Figure~\ref{fig:cbp_maps_q_full_da} shows the fractional semi-major axis variation, $\Delta a_p/a_{p,0}$, while Fig.~\ref{fig:cbp_maps_q_full_de} shows the eccentricity variation, $\Delta e_p$, both evaluated on the same $(f,e_{p,0})$ grid for surviving planets after the tidal forcing ends.
In each tile, the corresponding binary mass ratio $q_{\rm B}=m_2/m_1$ is indicated inside the panel, and unstable outcomes (ejections or stellar collisions) are marked in grey but not colour-coded.
Presenting the two quantities in separate galleries allows a cleaner comparison of the dynamical response across $q_{\rm B}$, while the unified colour scale within each figure makes it easier to identify systematic trends with binary mass ratio.
Overall, the maps show that the domain of long-lived survivors contracts as $q_{\rm B}$ increases, while the response of the surviving population becomes more structured and, in several regions of parameter space, more widely dispersed.

\section{Excess loss relative to fixed-binary controls}
\label{app:excess_loss}

The total fate budget discussed in Sect.~\ref{subsec:fates_vs_q} combines two contributions. 
Because our initial grid intentionally samples the near-boundary region, including configurations close to or slightly inside the initial stability limit, some planets are already short-lived in the corresponding fixed-binary problem. 
These early losses are part of the classical marginal-stability structure of circumbinary phase space. 
To isolate the additional contribution produced by binary evolution, we use the matched control integrations described in Sect.~\ref{sec:parspaceexpl}.

For each initial condition, we compare the outcome of the evolving-binary run with that of the corresponding fixed-binary control. 
We then identify the transition,
\begin{equation}
    S_{\rm ctrl}\rightarrow L_{\rm tides},
\end{equation}
where $S_{\rm ctrl}$ denotes a planet that survives in the control integration, and $L_{\rm tides}$ denotes a planet that is lost, either by ejection or Roche collision, when binary contraction and eccentricity damping are imposed. 
This transition provides a direct measure of the excess loss induced by the time-dependent binary potential, separated from losses already present in the static near-boundary problem.

\begin{figure}[H]
\centering

\begin{minipage}[t]{0.5\columnwidth}
    \vspace{0pt}
    \centering
    \includegraphics[width=0.75\linewidth]{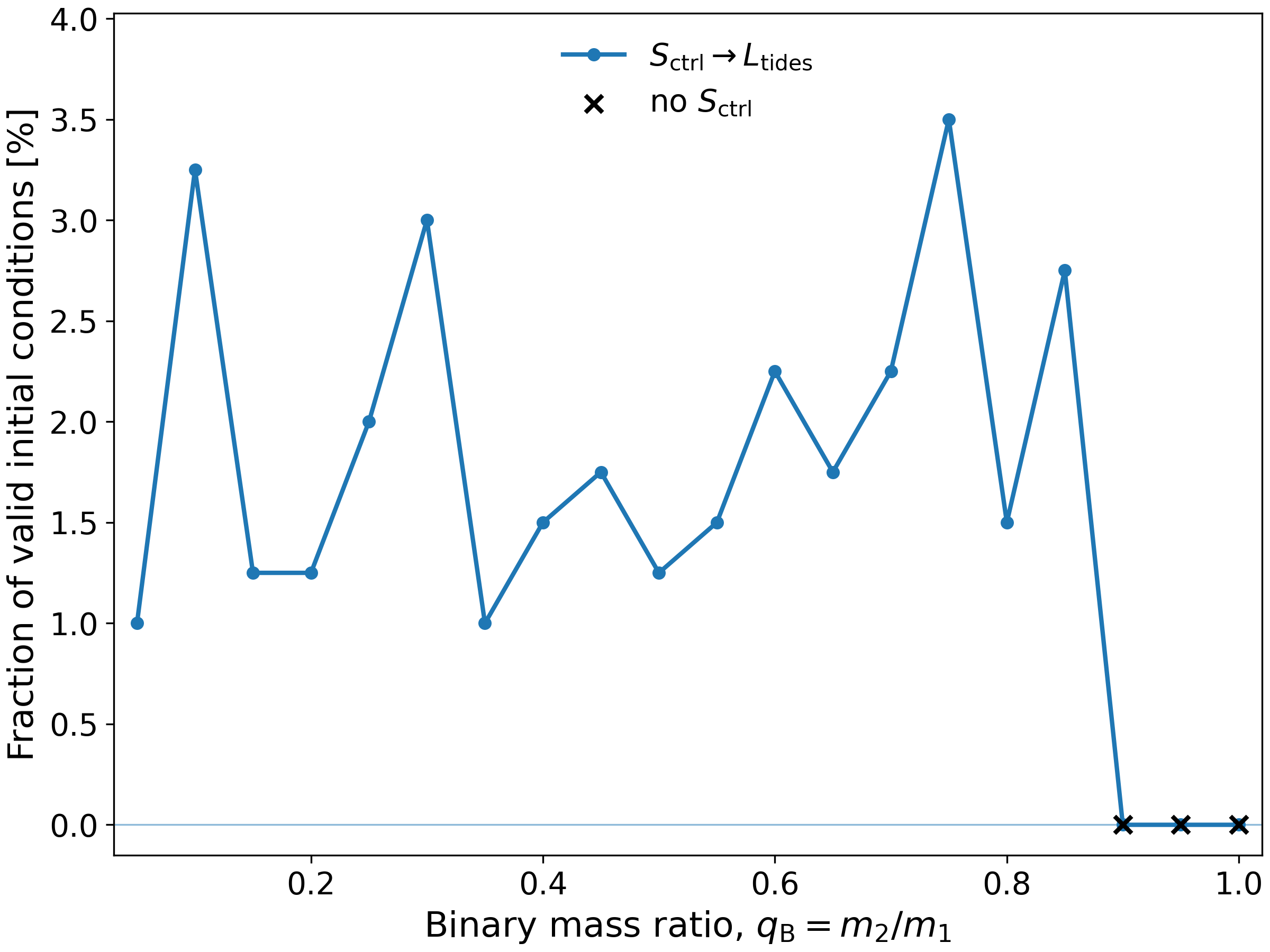}
\end{minipage}
\hfill
\begin{minipage}[t]{0.49\columnwidth}
    \vspace{0pt}
    \captionsetup{aboveskip=0pt}
    \captionof{figure}{
    Excess planet loss induced by binary evolution.
    For each binary mass ratio, we compare the evolving-binary integrations with their matched fixed-binary controls and identify planets that survive in the control run but are lost when binary contraction and eccentricity damping are imposed. The plotted fraction measures the transition
    $S_{\rm ctrl}\rightarrow L_{\rm tides}$ over the full set of valid initial conditions. In the fiducial experiment, all such induced losses occur through ejection and take place during the forcing interval. This component is modest compared with the total loss budget, which is dominated by initially marginal configurations, but it isolates the additional filtering produced by the moving stability boundary.
    }
    \label{fig:excess_loss}
\end{minipage}
\end{figure}

Figure~\ref{fig:excess_loss} shows this excess-loss fraction as a function of binary mass ratio. 
The fraction is computed over the full set of valid initial conditions at each $q_{\rm B}$. 
In the fiducial experiment, the induced component is modest, reaching a few per cent of the sampled grid. 
However, it is a clean diagnostic of the moving-boundary mechanism: these are planets that would survive around a fixed binary but are removed once the binary evolves. 
All induced losses in this subset occur through ejection rather than Roche collision.

We also checked the timing of these induced losses. 
All $S_{\rm ctrl}\rightarrow L_{\rm tides}$ events occur during the forcing interval, rather than before the onset of binary evolution or after the forcing has ended. 
This confirms that they are temporally associated with the imposed retreat of $a_{\rm crit}(t)$ and with the corresponding deformation of the near-boundary phase space. 
Thus, although the global loss budget is dominated by initially marginal configurations, the matched-control comparison demonstrates the existence of an additional contraction-driven ejection channel.

\section{Sensitivity to planetary mass}
\label{app:planet_mass}

 {Our fiducial simulations adopt an Earth-mass planet, whereas several observed CBPs have Saturn-to-Jupiter masses. To test the sensitivity of our results to this choice, we repeated a reduced $12\times12$ grid for six representative binary mass ratios using $m_p=1\,M_{\rm J}$ and a Jupiter-like bulk density. The initial conditions and orbital phases were kept matched to the corresponding Earth-mass integrations, giving 864 matched pairs.}

 {The global outcome budget is nearly unchanged. The Earth-mass runs yield 175 survivors, 550 ejections, and 139 Roche losses, compared with 174, 550, and 140, respectively, for the Jupiter-mass runs. Only 11 out of 864 initial conditions ($1.27\%$) change between survival and loss. Among planets surviving in both experiments, the median paired change in $\Delta_f$ remains below $10^{-3}$ in absolute value for $q_B\leq0.7$.}

 {The main difference concerns the partition between loss channels (see Table \ref{tab:mass_test}). At $q_B=1$, the Roche-loss fraction increases from $41.0\%$ in the Earth-mass case to $47.9\%$ in the Jupiter-mass case, with a corresponding decrease in the ejection fraction. This is consistent with the larger Roche radius adopted for the lower-density giant planet. We therefore find that the survival statistics and final displacement of the surviving population are robust to the adopted planetary mass at the level tested here, while the relative contribution of ejections and Roche losses retains some dependence on planetary density.}

\begin{table}[H]
\caption{ {Outcome statistics for the reduced Earth- and
Jupiter-mass experiments.}}
\label{tab:mass_test}
\centering
\begin{tabular}{lccc}
\hline\hline
Planet & Survive & Eject & Roche loss \\
\hline
$1\,M_{\oplus}$ & 175 & 550 & 139 \\
$1\,M_{\rm J}$  & 174 & 550 & 140 \\
\hline
\end{tabular}
\end{table}

\section{Sensitivity to the binary-evolution timescale}
\label{app:timescale}

 {
The fiducial experiments adopt $\tau_e=\Delta T=100$ kyr as a computationally convenient secular timescale rather than as a physical tidal-circularisation time. To test whether the planetary response depends on this choice, we repeated a reduced $12\times12$ subset of the original grid for $\tau_e=\Delta T=30$ and $300$ kyr, using $q_{\rm B}=0.30$, $0.70$, and $1.00$. The same initial conditions and binary evolutionary track were used in all cases, and the fixed-binary controls were integrated for the corresponding total integration time.}

 {
The outcome fractions are only weakly affected by the forcing timescale.
For $q_{\rm B}=0.30$, the survival fractions are $39.6$, $36.1$, and $36.1$ per cent for $\tau_e=30$, $100$, and $300$ kyr, respectively;
for $q_{\rm B}=0.70$ they are $15.3$, $14.6$, and $13.9$ per cent. For $q_{\rm B}=1.00$, no planets survive at any of the three timescales. The collision fractions remain unchanged at $4.2$, $12.5$, and $41.0$ per cent for $q_{\rm B}=0.30$, $0.70$, and $1.00$, respectively, with the remaining losses occurring through ejection.}

 {
The survivor distribution shows a similarly modest dependence. The median
$\Delta_f$ changes from $1.811$ to $1.854$ and $1.857$ for
$q_{\rm B}=0.30$, and from $1.913$ to $1.951$ and $1.966$ for $q_{\rm B}=0.70$, as the forcing timescale increases from $30$ to $100$ and $300$ kyr. Relative to the fiducial case, only $3.5$ per cent of the initial conditions change between survival and loss in the most sensitive case ($q_{\rm B}=0.30$, $30$ kyr), decreasing to $1.4$ per cent for $300$ kyr and to $0.7$ per cent or less for $q_{\rm B}=0.70$ and $1.00$. Thus, faster forcing produces a small residual increase in survival and slightly smaller final offsets, but the survival/loss budget and the population-level $\Delta_f$ distribution remain qualitatively unchanged over the explored factor-of-ten range in timescale.}

\end{appendix}
\end{document}